\documentclass[12pt, a4paper]{article}
\usepackage{graphicx}
\usepackage{amssymb}
\usepackage{amsmath}
\usepackage{bm}
\usepackage{color}
\usepackage{theorem}
\usepackage{subcaption}
\usepackage{listings}
\usepackage{colortbl}
\usepackage{tabularx}
\usepackage{longtable}
\usepackage[utf8]{inputenc}
\usepackage[T1]{fontenc}
\usepackage{lmodern}
\usepackage[top=30truemm,bottom=30truemm,left=25truemm,right=25truemm]{geometry}
\usepackage[low-sup]{subdepth}
\usepackage{soul}

\usepackage[sort&compress,numbers, merge]{natbib}
\definecolor{gesfpurple}{rgb}{0.47,0.19,0.42}

\definecolor{gesflanse}{rgb}{0.00,0.50,0.50}

\definecolor{gesfblue}{rgb}{0.08,0.42,0.76}

\definecolor{gesfred}{rgb}{1,0,0}

\definecolor{gesfwhite}{rgb}{1,1,1}

\definecolor{gesfblack}{rgb}{0,0,0}

\definecolor{Orange}{cmyk}{0,0.61,0.87,0}
\definecolor{JungleGreen}{cmyk}{0.99,0,0.52,0}
\definecolor{OliveGreen}{cmyk}{0.64,0,0.95,0.40}
\definecolor{Brown}{cmyk}{0,0.81,1,0.60}
\definecolor{RoyalBlue}{cmyk}{0.71,0.53,0,0.12}
\definecolor{Gray}{cmyk}{0,0,0,0.40}
\definecolor{LightPink}{cmyk}{0.0,0.25,0,0}
\definecolor{LLightPink}{cmyk}{0.0,0.10,0,0}
\definecolor{LightBlue}{cmyk}{0.25,0,0,0}
\definecolor{LightGray}{cmyk}{0,0,0,0.2}

\newcommand{\dd}{{\rm d}}

\allowdisplaybreaks[1]

\usepackage[colorlinks=true, linkcolor=OliveGreen, citecolor=RoyalBlue,
urlcolor=RoyalBlue]{hyperref}

\renewcommand{\thefootnote}{\fnsymbol{footnote}}

\begin{document}

\begin{titlepage}

  \begin{flushright}
\end{flushright}

\vskip 1.35cm
\begin{center}

{\large
{\bf
Supernova-scope for the Direct Search of Supernova Axions
}
}

\vskip 1.5cm

Shao-Feng~Ge$^{a,b,c}$, 
Koichi~Hamaguchi$^{d,e}$,
Koichi~Ichimura$^{f,e}$
Koji~Ishidoshiro$^f$,
Yoshiki~Kanazawa$^d$, 
Yasuhiro~Kishimoto$^{f,e}$,
Natsumi~Nagata$^d$,
Jiaming~Zheng$^{a,b}$

\vskip 0.8cm

{\it $^a$Tsung-Dao Lee Institute, Shanghai 200240, Shanghai Jiao Tong University, China}\\[2pt]
{\it $^b$School of Physics and Astronomy,
 Shanghai Jiao Tong University, Shanghai 200240, China}\\[2pt]
{\it $^c$Shanghai Key Laboratory for Particle Physics and Cosmology, Shanghai Jiao Tong University, Shanghai 200240, China}\\[2pt]
{\it $^d$Department of Physics, University of Tokyo, Bunkyo-ku, Tokyo
 113--0033, Japan} \\[2pt]
{\it $^e$Kavli IPMU (WPI), University of Tokyo, Kashiwa, Chiba
 277--8583, Japan} \\[2pt]
{\it $^f$Research Center for Neutrino Science, Tohoku University, Sendai
 980--8578, Japan} 

\date{\today}

\vskip 1.5cm

\begin{abstract}

If a supernova explosion occurs within a few hundred {parsecs} from the Earth,
a huge number of axions, in addition to neutrinos, may arrive at the Earth.
In this paper, we discuss in detail the prospect of detecting those supernova axions by an axion helioscope.
With the help of a pre-supernova neutrino
alert system, it is possible to point a helioscope 
at an exploding supernova in advance.
The supernova axions can then be detected by a gamma-ray detector installed at the end of the helioscope.
We call such a detection system an {\it axion supernova-scope} (SNscope).
We propose a conceptual design for an axion SNscope, where the gamma-ray detector is installed at the opposite end to the X-ray detector for the solar axion. 
It still functions as an axion helioscope during the normal operation time, and once a pre-SN neutrino alert is received, the scope is temporarily turned around and targeted to a SN candidate, waiting for the supernova axions.
We estimate the sensitivity of supernova axion detection and find that SNscopes based on the next-generation axion helioscopes, such as IAXO, have potential to explore the invisible axions and to test the axion interpretation of stellar cooling anomalies.

\end{abstract}

\end{center}
\end{titlepage}

\renewcommand{\thefootnote}{\arabic{footnote}}
\setcounter{footnote}{0}

\section{Introduction}

Axion~\cite{Weinberg:1977ma, Wilczek:1977pj} is a pseudo Nambu-Goldstone
boson associated with the spontaneous breaking of the Peccei-Quinn
symmetry~\cite{Peccei:1977hh, Peccei:1977ur}, which was introduced to
solve the strong CP problem. Although the original axion model has
already been excluded by experiment, its simple extensions---called
invisible axion models~\cite{Kim:1979if, Shifman:1979if,
Zhitnitsky:1980tq, Dine:1981rt}---are perfectly viable
and capable of solving
the strong CP problem. The characteristic features of the invisible
axions are tiny mass and extremely weak couplings to the Standard Model
particles.
The latter property helps the invisible axions
to evade the experimental constraints with the
price of making it very
challenging to test these models.

Stellar objects offer promising ways
of testing axions and provide
stringent constraints on axion models.
For instance, the observed duration of
the neutrino signal from SN1987A \cite{Hirata:1987hu, 
Bionta:1987qt, Alekseev:1987ej} provides one of the strongest bounds on
axions~\cite{Ellis:1987pk, Raffelt:1987yt, Turner:1987by, Mayle:1987as, 
Brinkmann:1988vi, Burrows:1988ah, Mayle:1989yx, Chang:2018rso,
Carenza:2019pxu}.\footnote{It is, however, pointed out that this bound may not be
robust~\cite{Bar:2019ifz}, given that the present understanding of
SN1987A is limited.} The temperature observations of certain neutron
stars also give severe limits on axion models~\cite{Hamaguchi:2018oqw,
Beznogov:2018fda, Leinson:2019cqv}, which
are as strong as the
SN1987A bound. On the other hand, several observations indicate
some preference for the presence of 
extra stellar energy
losses, which may be attributed to axion~\cite{Raffelt:2011ft, Giannotti:2015kwo, Giannotti:2017hny, Saikawa:2019lng}.

These astrophysical limits/hints are in general given by indirect searches of axions and suffer from considerable uncertainty from both theory and observation.
It would be of great advantages to have
direct searches for astrophysical axions.
The axion helioscopes
\cite{Sikivie:1983ip}
designed for detecting solar axions
are suitable and promising tools for this purpose. Several 
axion helioscope experiments have been performed so far and imposed
limits on the axion-photon coupling \cite{Lazarus:1992ry,
Moriyama:1998kd, Inoue:2002qy, Inoue:2008zp, Zioutas:2004hi,
Andriamonje:2007ew, Arik:2008mq, Arik:2011rx, Arik:2013nya,
Anastassopoulos:2017ftl}. Next-generation helioscopes, such as the
International Axion Observatory (IAXO) \cite{Armengaud:2014gea,
Armengaud:2019uso} and Trioitsk Axion Solar Telescope Experiment
(TASTE) \cite{Anastassopoulos:2017kag}, are being planned and expected
to have significantly improved sensitivities.

Axion helioscopes can be pointed at not only the Sun but also other
celestial objects. An interesting possibility of such an application is
to detect a burst of axions from a nearby supernova (SN) explosion. It is in
principle possible to target helioscopes at an exploding SN in advance
with the help of a pre-SN neutrino alert
system, such as the Supernova Early
Warning System (SNEWS)~\cite{Antonioli:2004zb}.
The SN axions then convert to photons
inside the helioscope, which
can be detected if a $\gamma$-ray detector is installed on the end of the helioscope.
Indeed, there have been brief speculations on such a possibility in
the literature \cite{Raffelt:2011ft, Irastorza:2018dyq}. 
In this paper, we study in detail the prospect of detecting SN axions with an operating helioscope and propose a conceptual design for the detection setup, which we call the \textit{axion supernova-scope (SNscope)}.
The gamma-ray detector is installed at the opposite end to the X-ray detector for the solar axion. 
The experiment can work as an axion helioscope during the normal operation time, and once a pre-SN neutrino alert is received, the scope is turned around and targeted at a SN candidate, waiting for the SN axions.
We show that 
the axion SNscopes based on 
the next-generation axion helioscopes can probe invisible axions 
if a SN explosion occurs within a few hundred {parsecs} from the Earth.
The SN axion detection
is a realistic and promising option for
extending the physics potential of the
future axion helioscopes.

The outline of this paper is the following. In Sec.~\ref{sec:axion}, we
review relevant features of invisible axions and the present constraints on them. In Sec.~\ref{sec:nearbysn}, we list nearby SN progenitor candidates and discuss the pre-SN neutrino alert system that may forecast SN explosions. In Sec.~\ref{sec:layout}, we show the layout of axion SNscopes which we discuss in this paper. We then evaluate the observational probability for each SN progenitor candidate, estimate the number of SN axion events detected by an axion SNscope, and discuss the background for this in Sec.~\ref{sec:prospects}. Section~\ref{sec:conclusion} summarizes our conclusions.

\section{Axion}
\label{sec:axion}

Axion\footnote{For recent reviews on axion and the current status of
axion searches, see Refs.~\cite{Tanabashi:2018oca, Irastorza:2018dyq, Armengaud:2019uso, DiLuzio:2020wdo, Sikivie:2020zpn}.} \cite{Weinberg:1977ma,
Wilczek:1977pj} is a pseudo Nambu-Goldstone boson which appears below
the symmetry-breaking scale of the global U(1) Peccei-Quinn symmetry
\cite{Peccei:1977hh, Peccei:1977ur}. This 
energy scale is characterized by 
the axion decay constant $f_a$.
For the invisible axion, such as the
KSVZ~\cite{Kim:1979if, Shifman:1979if} and
DFSZ~\cite{Zhitnitsky:1980tq, Dine:1981rt} models,
$f_a$ is
much larger than the
electroweak scale and the interactions of the axion
at low energies are
described by the following effective Lagrangian,
\begin{align}
 {\cal L}_{\rm int} &=  \frac{\alpha_s}{8\pi} \frac{a}{f_a} G^{a\mu\nu}
 \widetilde{G}^a_{\mu\nu} 
+ \frac{g_{a\gamma\gamma}}{4} a F_{\mu\nu} \widetilde{F}^{\mu\nu} 
+ \sum_{f} \frac{C_f}{2f_a} \bar{f} \gamma^\mu \gamma_5 f \partial_\mu a
 + \dots ~,
\label{eq:lag}
\end{align}
where $a$ is the axion field, $\alpha_s \equiv g_s^2/(4\pi)$ with
$g_s$ being the strong gauge coupling constant,
$G^a_{\mu\nu}$ and
$F_{\mu\nu}$ are the field strength tensors of the color and
electromagnetic gauge fields, respectively, $\widetilde{G}^a_{\mu\nu}
\equiv \frac{1}{2} \epsilon_{\mu\nu\rho\sigma} G^{a\rho \sigma}$ and
$\widetilde{F}_{\mu\nu} \equiv \frac{1}{2} \epsilon_{\mu\nu\rho\sigma}
F^{\rho\sigma}$ with $\epsilon^{\mu\nu\rho\sigma}$ the totally
antisymmetric tensor,
$f$ denotes the
Standard Model fermions, and
the dots indicate higher-dimensional operators which are irrelevant to our
discussions. 

At the leading order in chiral perturbation theory with two quark
flavors, the axion mass $m_a$ is~\cite{Weinberg:1977ma}
\begin{equation}
 m_a = \frac{\sqrt{m_u m_d}}{m_u + m_d} \frac{f_\pi m_\pi}{f_a}
\simeq 5.8 \times \biggl(\frac{f_a}{10^9~\mathrm{GeV}}\biggr)^{-1}
~\mathrm{meV}~,
\label{eq:ma}
\end{equation}
where $m_u = 2.16$~MeV, $m_d = 4.67$~MeV, $m_\pi = 135$~MeV are the
masses of up quark, down quark, and the neutral pion, respectively, and
$f_\pi = 92.1$~MeV~\cite{Tanabashi:2018oca} is the pion decay
constant. The next-to-next-to-leading order
computation in chiral perturbation theory and a recent calculation with
QCD lattice simulations of $m_a$ can be found in
\cite{Gorghetto:2018ocs} and \cite{Borsanyi:2016ksw}, respectively,
whose results are consistent with Eq.~\eqref{eq:ma}
up to ${\cal O}(1-10)\%$ corrections.

The coupling of the axion-photon interaction, $g_{a\gamma\gamma}$, is a
model-dependent parameter. 
Its value is inversely proportional to $f_a$ and is given by, at the leading order in chiral
perturbation theory,
\begin{equation}
 g_{a\gamma\gamma} = \frac{\alpha}{2\pi f_a} \biggl[
\frac{E}{N} - \frac{2}{3}\frac{4 m_d + m_u}{m_u + m_d}
\biggr]
\simeq  \frac{\alpha}{2\pi f_a} \biggl[
\frac{E}{N} - 2.0
\biggr] ~,
\label{eq:gagg}
\end{equation}
where $\alpha$ is the fine-structure constant, and $E/N$ is the ratio
between the electromagnetic and color anomaly factors for the
Peccei-Quinn current; for instance, $E/N = 8/3$ in the DFSZ model while
$E/N = 0$ in the KSVZ model with electrically neutral Peccei-Quinn
fermions. The next-to-leading order calculation of $g_{a\gamma\gamma}$
in chiral perturbation theory is available in
\cite{diCortona:2015ldu}, which is in good agreement with
Eq.~\eqref{eq:gagg}. This axion-photon coupling determines the
axion-to-photon conversion rate in a helioscope/SNscope.

At low energies, the axion-gluon (the first term) and axion-quark (the
third term) interactions in Eq.~\eqref{eq:lag} induce the axion-nucleon
couplings, which have the form
\begin{equation}
  {\cal L}_{aNN}
=
  \sum_{N =p,n} \frac{C_N}{2 f_a}
  \bar{N} \gamma^\mu \gamma_5 N \partial_\mu a
\equiv
  \sum_{N = p,n}
  \frac {g_{aN}}{2 m_N}
  \bar{N} \gamma^\mu \gamma_5 N \partial_\mu a,
\label{eq:axion_nucleon}
\end{equation}
where $g_{aN} \equiv C_{N} m_N / f_a$.
At the leading order in $\alpha_s$, the coefficients $C_N$ are given by 
\begin{equation}
 C_N = \sum_{q} \biggl(C_q - \frac{m_*}{m_q}\biggr)\Delta q^{(N)} ~,
\label{eq:cn}
\end{equation}
where $m_* \equiv m_um_dm_s/(m_u m_d + m_d m_s + m_u m_s)$ with $m_s$ being
the strange quark mass.
The factor $\Delta q^{(N)}$ is the spin fraction
defined by $2 s_\mu^{(N)} \Delta q^{(N)} \equiv \langle N| \bar{q}
\gamma_\mu \gamma_5 q |N\rangle$ with $s_\mu^{(N)}$
denoting the spin of the
nucleon $N$: $\Delta u^{(p)} = \Delta d^{(n)} =
0.897(27)$, $\Delta d^{(p)} = \Delta u^{(n)} = -0.376(27)$, and $\Delta
s^{(p)} = \Delta s^{(n)} = -0.026(4)$ \cite{diCortona:2015ldu}. The QCD
corrections to Eq.~\eqref{eq:cn} are considered in
\cite{diCortona:2015ldu}. For the KSVZ axion ($C_q = 0$), we have
\cite{diCortona:2015ldu} 
\begin{equation}
 C_p = -0.47(3)~, \qquad C_n =-0.02(3) ~, 
\end{equation}
and for the DFSZ axion ($C_{u,c,t} = \cos^2 \beta/3$ and
$C_{d,s,b} = \sin^2 \beta/3$ with $\tan\beta$
denoting the ratio of the vacuum
expectation values of the two doublet Higgs fields, $\tan \beta \equiv
\langle H_u\rangle /\langle H_d\rangle$)~\cite{Tanabashi:2018oca},
\begin{subequations}
\begin{align}
 C_p & = -0.182(25) - 0.435 \sin^2 \beta~, \\
 C_n & = -0.160(25) + 0.414 \sin^2 \beta~.
\end{align}
\end{subequations}
These axion-nucleon couplings determine the production rate of axions in
SN explosions.

Let us now discuss the limits on the  
axion couplings from stellar cooling.
One of
the strongest constraints is derived
from the measured neutrino signal duration
of SN1987A \cite{Ellis:1987pk, Raffelt:1987yt,
Turner:1987by, Mayle:1987as, 
Brinkmann:1988vi, Burrows:1988ah, Mayle:1989yx,
Chang:2018rso, Carenza:2019pxu}. 
In a recent analysis~\cite{Carenza:2019pxu}, 
the constraint on the axion-nucleon couplings is obtained as 
$g_{an}^2+0.61g_{ap}^2+0.53 g_{an}g_{ap}\lesssim 8.26\times 10^{-19}$, 
by imposing the condition
that the axion luminosity ($L_a$) at the post-bounce time $t_{\rm pb}=1$ sec 
should 
not exceed the neutrino luminosity ($L_\nu$),
$L_a\lesssim L_\nu \simeq 2\times 10^{52}\,{\rm erg}\;{\rm s}^{-1}$.
This corresponds to $f_a\gtrsim 3.9\times 10^8$\,GeV for KSVZ and $f_a\gtrsim 4.7\times 10^8$\,GeV for DFSZ with $\tan \beta = 10$. 
We note that these constraints should be
considered as indicators rather than sharp
bounds with the present limited understanding of the
SNe \cite{Tanabashi:2018oca}. 
It was also pointed out recently~\cite{Bar:2019ifz} that
the observed neutrinos from SN1987A might have come from an accretion disk that would not be cooled by emitting axions. If this is the case, the criteria $L_a\lesssim L_\nu$ is not applicable and neither are the bounds obtained with it.

Another stringent constraint on the axions is given
by the temperature observations of neutron
stars~\cite{Hamaguchi:2018oqw, Beznogov:2018fda,
Leinson:2019cqv}. The cooling of the neutron star
in the SN remnant Cassiopeia A (Cas A) is studied in Ref.~\cite{Hamaguchi:2018oqw}, which gives $f_a\gtrsim \sqrt{0.9 C_p^2 + 1.4 C_n^2}\times 10^9$~GeV; 
this leads to $f_a \gtrsim 5 \times 10^8$~GeV
for KSVZ and $f_a \gtrsim 7 \times 10^{8}$~GeV for
DFSZ with $\tan \beta = 10$. 
Another hot young
neutron star in the SN remnant HESS J1731-347
requires $f_a\gtrsim 3.4 |C_n| \times 10^9$~GeV
\cite{Beznogov:2018fda}, which gives $f_a \gtrsim 7
\times 10^7$~GeV for KSVZ and $f_a \gtrsim 8 \times
10^8$~GeV for DFSZ with $\tan \beta = 10$.
We stress again that there can be ${\cal O}(1)$ uncertainty in these bounds (see, \textit{e.g.}, the discussion on the uncertainty coming from the envelope composition of the Cas A neutron star in 
\cite{Hamaguchi:2018oqw}).
For the other stellar cooling constraints on the axion couplings, see~\cite{Tanabashi:2018oca} and references therein, and also \cite{Capozzi:2020cbu}.

Finally, it has been pointed out that several observations of other astrophysical objects, such as white dwarfs, red giant branch stars, and horizontal branch stars, may point to the existence of additional source of stellar cooling beyond the standard cooling sources, and it may be the hint of the axion~\cite{Raffelt:2011ft, Giannotti:2015kwo, Giannotti:2017hny, Saikawa:2019lng}. This cooling hint prefers a value of $f_a$ lower than the limits quoted above; for example, in \cite{Giannotti:2017hny}, the best fit value for $f_a$ is obtained for the DFSZ model as $f_a = 7.7 \times 10^7$~GeV and $\tan\beta=0.28$ if the SN1987A constraint is not included. These cooling hints also suffer from large uncertainty in both theory and observation.

Given the significance of these astrophysical
limits/hints on axions as well as their large uncertainty, 
it is desirable to consider a more
direct way of probing axions produced from the
stellar objects mentioned above. Among them, SNe are the most promising target because of the prominent luminosity of axions emitted during the first ten seconds from the explosion. In what follows, we discuss a strategy for the detection of SN axions
by adapting an axion helioscope\footnote{
Other aspects of detecting SN axions are also discussed in the literature, such as the detection of SN axions at Hyper-Kamiokande in \cite{Carenza:2018jjc, Carenza:2018vcb}, the prospect of probing diffusive SN axion background in \cite{Raffelt:2011ft} and the possible observation using Fermi Large Area Telescope in \cite{Meyer:2016wrm,Meyer:2020vzy}.
}.

\section{Nearby SN candidates and pre-SN neutrino alarm}
\label{sec:nearbysn}

Even before a core-collapse SN explosion, a large number of neutrinos are
emitted from the progenitor~\cite{Odrzywolek:2003vn}. These pre-SN neutrinos are produced during the last stages of the stellar evolution and typically have energies of 1--2~MeV. The calculation of the flux of pre-SN neutrinos has been performed in the literature~\cite{Odrzywolek:2003vn, Odrzywolek:2004em, 2009AcPPB..40.3063K, Odrzywolek:2009wa, Kato:2015faa, Patton:2015sqt, Yoshida:2016imf, Kato:2017ehj,Patton:2017neq, Guo:2019orq, Kato:2020hlc}, showing that it is possible to detect them if the progenitor star is located sufficiently near the Earth ($\lesssim 1$~kpc). In particular, the on-going/future neutrino experiments, \textit{e.g.}, KamLAND~\cite{Asakura:2015bga}, SNO+~\cite{Andringa:2015tza}, Super-Kamiokande~\cite{Simpson:2019xwo}, Hyper-Kamiokande~\cite{Abe:2018uyc}, JUNO~\cite{An:2015jdp}, and DUNE~\cite{Abi:2018dnh, Abi:2020evt}, as well as future dark matter detectors~\cite{Raj:2019wpy}, are expected to be sensitive to pre-SN neutrinos, which provide a possibility of forecasting the occurrence of a SN event in advance~\cite{Odrzywolek:2003vn}.
This global network of these experiments developed for an early SN alarm is called SNEWS~\cite{Antonioli:2004zb, Scholberg:2008fa, Vigorito:2011zza}.\footnote{For a recent status of SNEWS, see \cite{Benzvi:2020}. } 

\begin{table}[t]
  \centering
  \caption{List of SN progenitor candidates with having a mass $\gtrsim 10~M_{\odot}$ and within $250$~pc from the Earth. We basically use the values listed in the Hipparcos catalogue~\cite{vanLeeuwen:2007tv}; otherwise, we show the reference for the source. }
  {\footnotesize
  \begin{tabular}{llllll}
    \hline\hline
    HIP &Common Name & Distance (pc)& Mass ($M_\odot$)& RA (J2000) &Dec (J2000) \\ 
    \hline
    65474& Spica/$\alpha$ Virginis & 77(4) & $11.43\pm 1.15$~\cite{2016MNRAS.458.1964T} &13:25:11.58 & $-11$:09:40.8 \\
    81377& $\zeta$ Ophiuchi & 112(3) & $20.0$~\cite{2001MNRAS.327..353H} &16:37:09.54 & $-10$:34:01.5 \\
    71860& $\alpha$ Lupi & 142(3) & $10.1\pm 1.0$~\cite{2011MNRAS.410..190T} &14:41:55.76 & $-47$:23:17.5 \\
    80763& Antares/$\alpha$ Scorpii &170(30) &11--14.3~\cite{2013AA...555A..24O}
    &16:29:24.46 &$-26$:25:55.2 \\
    107315&Enif/$\epsilon$ Pegasi &211(8) &11.7(8)~\cite{2011MNRAS.410..190T}&21:44:11.16 &$+09$:52:30.0 \\
    27989 & Betelgeuse/$\alpha$ Orionis & $222^{+48}_{-34}$~\cite{2017AJ....154...11H} &$11.6^{+5.0}_{-3.9}$~\cite{2011ASPC..451..117N} & 05:55:10.31 & $+07$:24:25.4 \\
    \hline\hline
  \end{tabular}
  }
  \label{tab:nearbystars}
\end{table}

\begin{figure}[t]
  \centering
  {\includegraphics[width=0.85\textwidth]{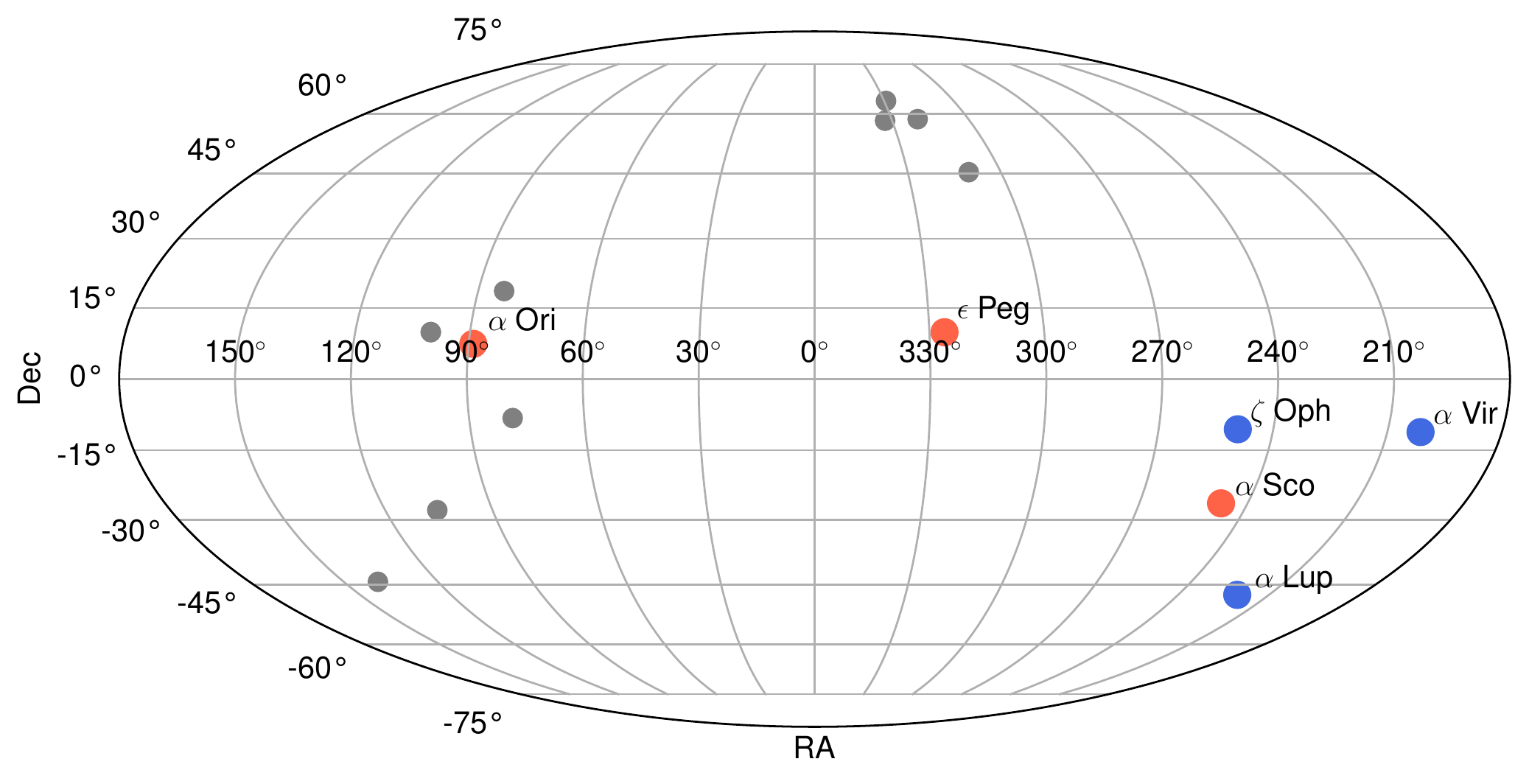}} 
  \caption{
    The position of the SN progenitors in Table~\ref{tab:nearbystars} on the Mollweide projection of the celestial sphere, where the red and blue dots correspond to the spectral types of K/M and O/B, respectively. We also show by the gray dots the progenitors with $d > 250$~pc and $M \gtrsim 10~M_{\odot}$ listed in Table~A1 in Ref.~\cite{Mukhopadhyay:2020ubs}.
  }
  \label{fig:distribution}
  \end{figure}

As we discuss in Sec.~\ref{sec:prospects}, SN axions may be detectable if the SN is within a few hundred parsecs from the Earth. For such a nearby SN, it is quite likely that the SNEWS gives an alert prior to the SN explosion. There are a sizable number of stars within this range that are considered to be exploding in the near future; we summarize the progenitor candidates of core-collapse SNe within the distance $d\leq 250$~pc from the Earth in Table~\ref{tab:nearbystars}, which we have taken from Refs.~\cite{Nakamura:2016kkl, nakamura, Mukhopadhyay:2020ubs}. In this table, we show both red and blue (super)giant stars\footnote{For instance, the progenitor of SN 1987A was found to be a blue supergiant~\cite{1987A&A...177L..25P, 1987Natur.328..318G, 1987ApJ...323L..35S}.
} that have a mass $M \gtrsim 10~M_{\odot}$,\footnote{We have not included those with a mass less than 10~$M_{\odot}$, since in this case it is found to be difficult to detect preSN neutrinos even if the progenitors are within 200~pc from the Earth~\cite{Kato:2015faa}. } for which an iron core is expected to be formed prior to the stellar collapse. 
For a more complete list of nearby SN progenitor candidates, see, \textit{e.g.}, Table~A1 of Ref.~\cite{Mukhopadhyay:2020ubs}. In  Table~\ref{tab:nearbystars}, the first column shows the Hipparcos Catalogue number, and the fifth and sixth columns are J2000 right ascension and declination, respectively. We also show the distribution of these stars on the Mollweide projection of the celestial sphere in Fig.~\ref{fig:distribution}, where the red and blue dots correspond to the spectral types of K/M and O/B, respectively. We also show by the gray dots the progenitors with $d > 250$~pc and $M \gtrsim 10~M_{\odot}$ listed in Table~A1 of Ref.~\cite{Mukhopadhyay:2020ubs}.

With pre-SN neutrinos, it is in principle possible to estimate the location of the source on the sky, which would be useful to identify the progenitor
\cite{Li:2020pvf, Mukhopadhyay:2020ubs}.
A liquid scintillator detector, such as JUNO~\cite{An:2015jdp}, can use the inverse
beta decay process, $\bar{\nu }_e + p \to n+e^+$, to reconstruct the direction of the incoming anti-neutrino. The angular resolution obtained with a JUNO-like detector at ${\cal O}(1)$~hours before the SN explosion is estimated to be $\simeq 60^\circ$~\cite{Mukhopadhyay:2020ubs} for a nearby progenitor as in Table~\ref{tab:nearbystars}. This information by itself can narrow the list of candidates down to a few.\footnote{Liquid scintillator detectors are sensitive also to pre-SN neutrinos coming from a progenitor farther than those in Table~\ref{tab:nearbystars}, up to distance $\lesssim 1$~kpc, such as those shown by the gray dots in Fig.~\ref{fig:distribution}. We may distinguish the former cases from the latter by estimating the distance to the progenitor from the event rate of pre-SN neutrinos. In any case, the axion detection is promising only for a nearby SN, and thus it would be a sensible strategy to target the axion SNscope at one of the close progenitor candidates such as those in Table~\ref{tab:nearbystars} when we receive a pre-SN neutrino alert. } Moreover, if a detector
with a better angular sensitivity is available in the future, the precision of the progenitor identification is considerably improved. For instance, a lithium-loaded liquid scintillator~\cite{2014NatSR...4E4708T} may be able to give an angular resolution of $\simeq 15^\circ$~\cite{Mukhopadhyay:2020ubs}, with which we can most likely identify the progenitor star uniquely. All in all, it is feasible to determine the exploding nearby progenitor with a pre-SN neutrino alert system by ${\cal O}(1)$ hours before the explosion, which thus makes it possible to direct an axion SNscope at the progenitor in advance.

\section{Axion SNscope}
\label{sec:layout}

An axion helioscope \cite{Sikivie:1983ip} is a tool to detect the axion flux from the Sun, in which an incoming axion is converted into a photon while it passes through the strong magnetic field inside the helioscope. The converted photon is then detected by a X-ray detector placed on its end side. Previous axion helioscope experiments have not yet detected axion signatures, imposing constraints on axion models~\cite{Lazarus:1992ry, Moriyama:1998kd, Inoue:2002qy, Inoue:2008zp, Zioutas:2004hi, Andriamonje:2007ew, Arik:2008mq, Arik:2011rx, Arik:2013nya, Anastassopoulos:2017ftl}. Among these experiments, the CERN Axion Solar Telescope (CAST) experiment gives the most stringent limit on the axion-photon coupling: $g_{a\gamma\gamma } < 0.66 \times 10^{-10}~{\rm GeV}^{-1}$ for $m_a < 0.02$~eV~\cite{Anastassopoulos:2017ftl}. 

\begin{table}[t]
  \centering
  \caption{List of the on-going and next-generation helioscopes as well as their (proposed) site, magnetic field strength $B$, length $L$, and cross sectional area $A$. }
  {
  \begin{tabular}{lllll}
    \hline\hline
    Experiment & (Proposed) site & $B$ (T) & $L$ (m) & $A$ (m$^2$) \\ 
    \hline
    CAST~\cite{Zioutas:2004hi, Andriamonje:2007ew, Arik:2008mq, Arik:2011rx, Arik:2013nya, Anastassopoulos:2017ftl} & CERN &9 & 9.3 & $2.9 \times 10^{-3}$ \\
    BabyIAXO~\cite{Armengaud:2019uso} & DESY& $\sim 2$ & 10 &$0.77$\\
    IAXO baseline~\cite{Armengaud:2014gea, Armengaud:2019uso}& DESY & $\sim 2.5$ & $20$ & $2.3$\\
    IAXO+~\cite{Armengaud:2019uso} & DESY & $\sim 3.5$ & $22$ & $3.9$ \\
    TASTE~\cite{Anastassopoulos:2017kag} & INR & 3.5 & 12 & 0.28 \\
    \hline\hline
  \end{tabular}
  }
  \label{tab:helioscopes}
\end{table}

There are several proposals for next-generation helioscopes, which we summarize with their properties in Table~\ref{tab:helioscopes}. IAXO~\cite{Armengaud:2014gea, Armengaud:2019uso} was proposed as a follow-up of CAST and is planned to be built at DESY. It consists of eight bores of 600~mm diameter and a 25~m long toroidal superconducting magnet, which provides $\sim 2.5$~T magnetic field in the bores on average. This setup, called the IAXO baseline, realizes a total cross section area of $2.3$~m$^2$. Prior to IAXO, an experiment with a prototype helioscope with two 10~m long bores, called BabyIAXO~\cite{Armengaud:2019uso}, is going to be performed. This prototype experiment offers not only physics outcome with a sensitivity better than that of CAST but also opportunities to test potential improvements in the technical design. With this feedback, we may have an improved version of IAXO in the future, which we refer to as the IAXO+~\cite{Armengaud:2019uso} in Table~\ref{tab:helioscopes}. Meanwhile, another axion helioscope called TASTE~\cite{Anastassopoulos:2017kag} is projected to be constructed in the Institute for Nuclear Research (INR) in Russia. We show the relevant parameters of these axion helioscopes---the magnetic field strength $B$, length $L$, and cross sectional area $A$---in Table~\ref{tab:helioscopes}.

\begin{figure}
  \centering
  \subcaptionbox{\label{fig:detector_norm}
  Axion helioscope
  \vspace{10mm}
  }
  {\includegraphics[width=0.9\textwidth]{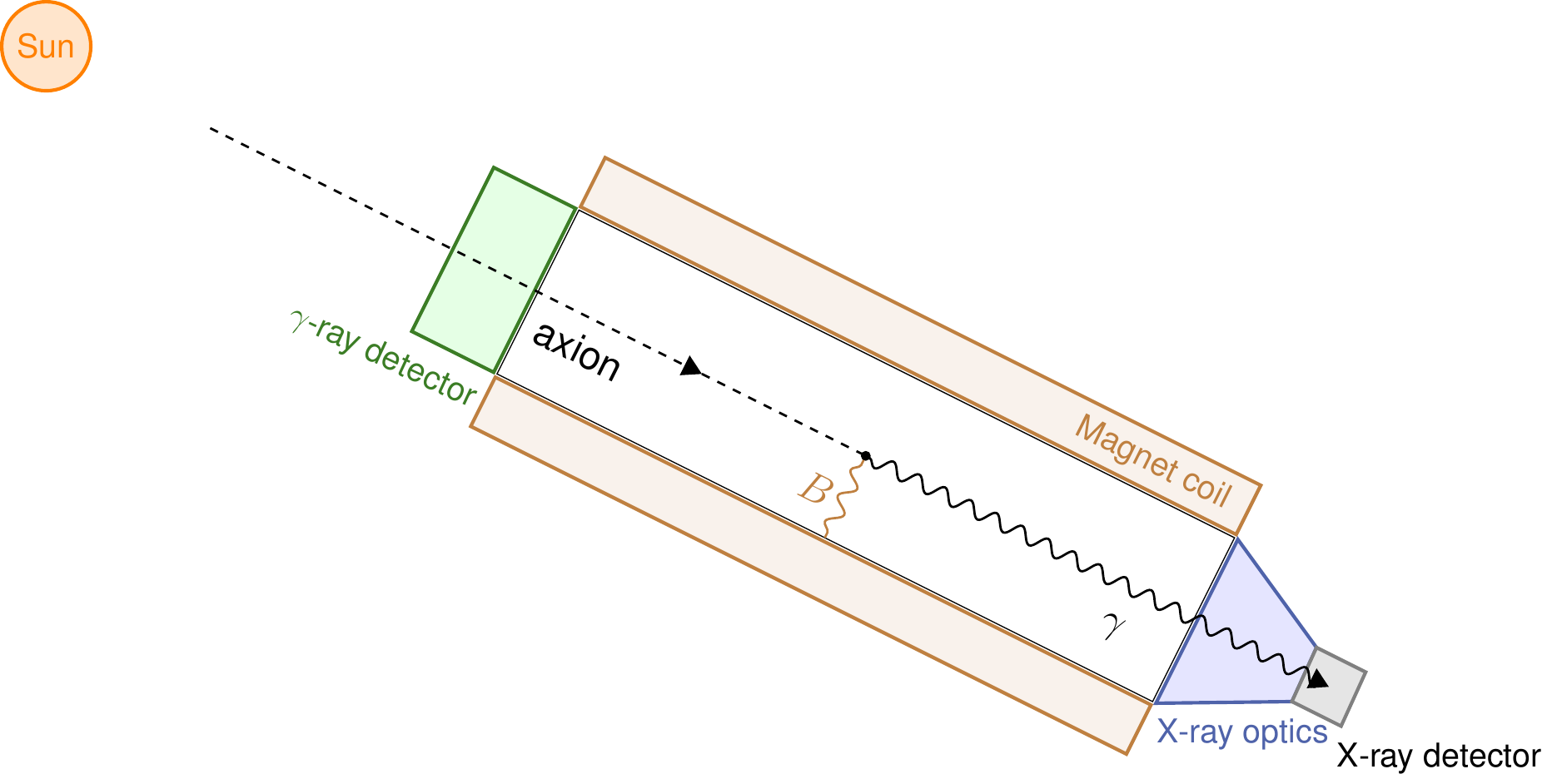}}
  \subcaptionbox{\label{fig:detector_preSN}
  Axion SNscope
  }
  { 
  \includegraphics[width=0.8\textwidth]{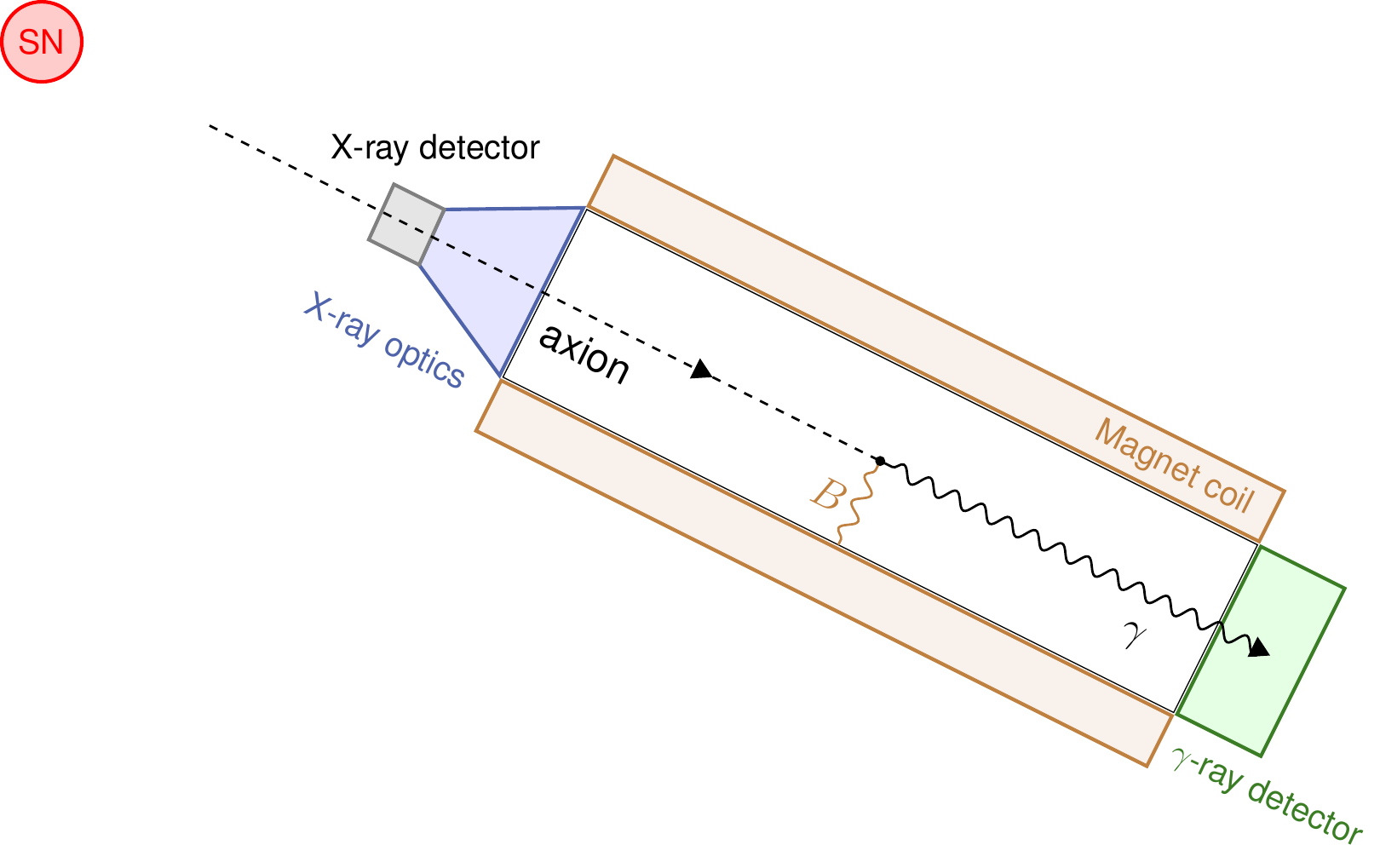}}
  \caption{
    Conceptual design of the axion detector discussed in this paper and its orientation for the detection of (a) solar and (b) SN axions. 
  }
  \label{fig:detector}
  \end{figure}

With the help of the pre-SN neutrino alarm discussed in the previous section, we can direct an axion helioscope at an exploding progenitor in advance, aiming at the detection of SN axions. Nevertheless, it is not possible to merely use an axion helioscope as it is for this purpose, since SN axions typically have energies of ${\cal O}$(10--100)~MeV
{as we will show in Sec.~\ref{sec:eventnumber}}
and thus the converted photons have a wavelength in the $\gamma $-ray range. 
This requires an additional installation of a $\gamma$-ray detector on a helioscope. For instance, we can put a $\gamma$-ray detector on the end side of a helioscope opposite to the X-ray detector, as illustrated in Fig.~\ref{fig:detector}. The conceptual design shown in this figure is basically the same as that for IAXO and TASTE, except for the additionally installed $\gamma$-ray detector indicated by the green box. We refer to this type of SN axion detectors as {axion SNscopes}. During the normal operation time, the $\gamma$-ray detector side is headed toward the Sun, as shown in Fig.~\ref{fig:detector_norm}, so that the converted photons are to be detected by the X-ray detector. When we receive a pre-SN neutrino alert, we turn the helioscope around to head the X-ray detector side toward a SN progenitor candidate, as shown in Fig.~\ref{fig:detector_preSN}. This setup allows us to use a solar axion helioscope for the SN axion detection as well, with a minimum modification. An example for the design of the $\gamma$-ray detector is given in Sec.~\ref{sec:bg}. In the next section, we discuss the prospects of such a SNscope for detecting axions from nearby SN progenitor candidates.

\section{Prospects}
\label{sec:prospects}

Now we discuss the feasibility of the SN axion detection with the setup discussed in the previous section. We first estimate the total fraction of time during which we can target a SNscope at progenitor stars in Sec.~\ref{sec:timefraction}. We then evaluate the expected number of signal and background events in Sec.~\ref{sec:eventnumber} and Sec.~\ref{sec:bg}, respectively, and discuss the prospects for the detection of SN axions.

\subsection{Observation time fraction}
\label{sec:timefraction}

Suppose an axion SNscope whose position in the equatorial coordinate system at time $t$ is $(\alpha_{\rm det}(t), \delta_{\rm det})$; for instance,  $\delta_{\rm det} \simeq 53.6^\circ$ for DESY, $\delta_{\rm det} \simeq 46.2^\circ$ for CERN, and $\delta_{\rm det} \simeq 55.5^\circ$ for INR. The azimuthal angle of the detector position, $\alpha_{\rm det} (t)$, varies as a linear function of time according to the Earth's rotation; the following discussion does not depend on its initial value and thus it will be set to a certain value for our convenience.  We assume that this SNscope can rotate by $360^\circ$ in the horizontal plane with the maximum elevation $\pm \theta_{\rm max}$; the planned values of $\theta_{\rm max}$ for IAXO and TASTE are $25^\circ$~\cite{Armengaud:2014gea} and $20^\circ$~\cite{Anastassopoulos:2017kag}, respectively. We use $\delta_{\rm det} \simeq 53.6^\circ$ and $\theta_{\rm max} = 25^\circ$, corresponding to IAXO, as representative parameters in the following analysis.

\begin{figure}
  \centering
  {\includegraphics[width=0.5\textwidth]{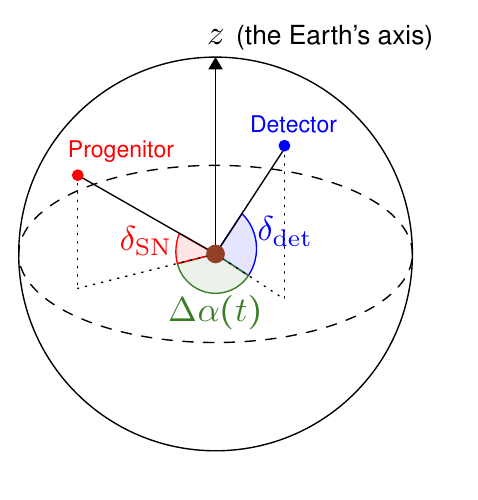}} 
  \caption{
    The positions of the detector (blue dot) and the target progenitor (red dot)  in our equatorial coordinate system. 
  }
  \label{fig:sphere}
  \end{figure}

We then consider a progenitor whose right ascension and declination are $\alpha_{\rm SN}$ and $\delta_{\rm SN}$, respectively, and define $\Delta \alpha (t) \equiv \alpha_{\rm det} (t) - \alpha_{\rm SN}$, as shown in Fig.~\ref{fig:sphere}. In the tangent plane at the detector position, the altitude of the progenitor, $\theta_{\rm SN} (t)$, is given by 
\begin{equation}
  \sin \theta_{\rm SN} (t) = \cos \delta_{\rm SN} \cos \delta_{\rm det} \cos \Delta \alpha (t) + \sin \delta_{\rm SN} \sin \delta_{\rm det} ~.
  \label{eq:sinthsn}
\end{equation}
If $|\theta_{\rm SN}| < \theta_{\rm max} $, we can target the SNscope at the progenitor.\footnote{
The finite size of the SNscope caliber may slightly increase the range of $|\theta_{\rm SN}|$.
For instance, the IAXO configuration may allow a field of view of roughly $0.6\,\mbox{m}/20\,\mbox{m} \approx 1.7^\circ$ as a SNscope.}

\begin{figure}
  \centering
  {\includegraphics[width=0.9\textwidth]{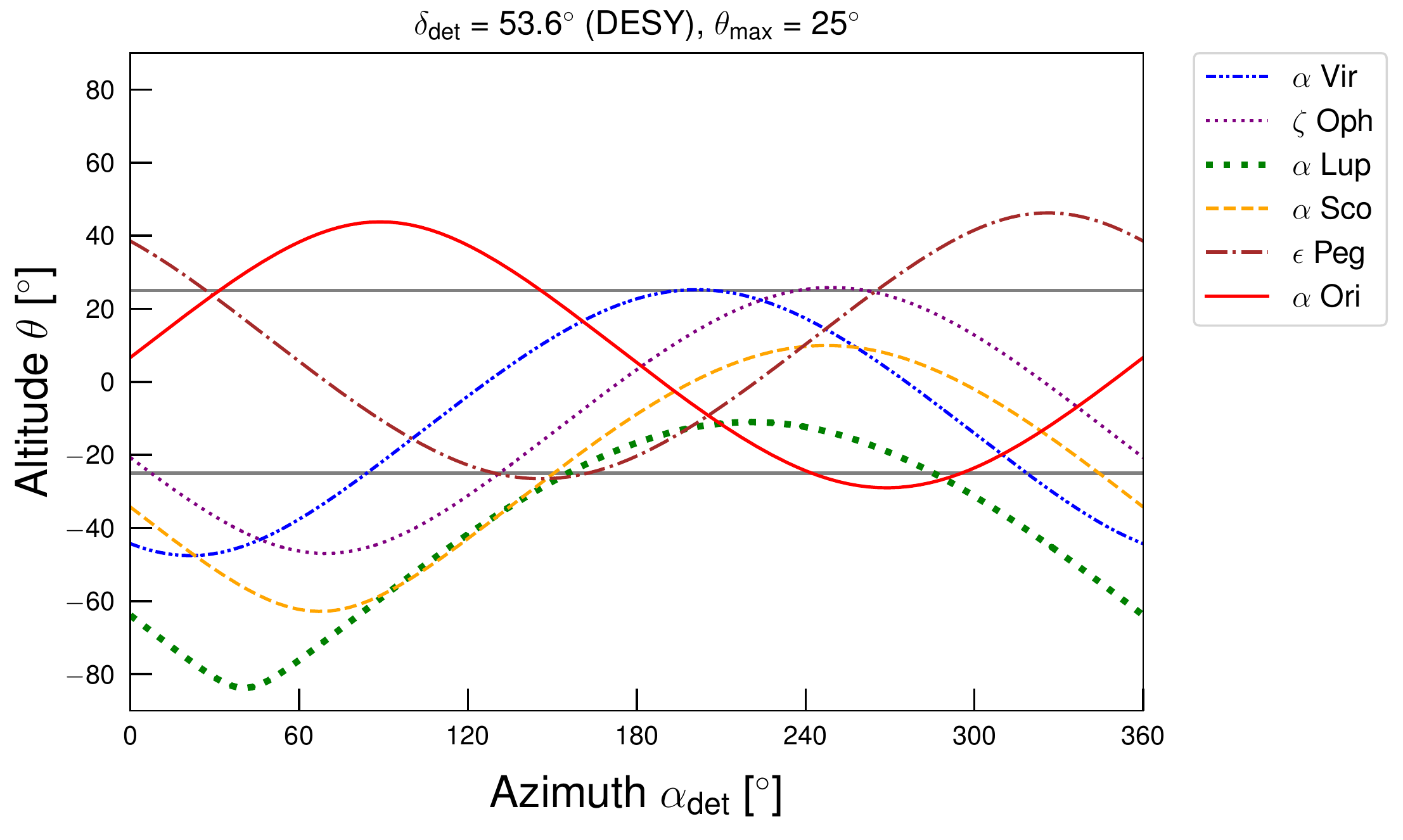}} 
  \caption{
    The altitude of the progenitors in Table~\ref{tab:nearbystars} observed at the detector position $(\alpha_{\rm det}, \delta_{\rm det})$ with $\delta_{\rm det} = 53.6^\circ$ (corresponding to DESY) as functions of $\alpha_{\rm det}$. The horizontal lines correspond to $\theta = \pm \theta_{\rm max}$ with $\theta_{\rm max} = 25^\circ$. 
  }
  \label{fig:altitude}
  \end{figure}

In Fig.~\ref{fig:altitude}, we show the altitude of the progenitors in Table~\ref{tab:nearbystars} observed at the detector position $(\alpha_{\rm det}, \delta_{\rm det})$ with $\delta_{\rm det} = 53.6^\circ$ (corresponding to DESY) as functions of $\alpha_{\rm det}$. The horizontal lines correspond to $\theta = \pm \theta_{\rm max}$ with $\theta_{\rm max} = 25^\circ$. 
From this figure, we see that the observational time fraction $\epsilon_t$ is larger than 50\% for all of the progenitors in Table~\ref{tab:nearbystars} except for $\alpha$ Lupi (green dotted line).

\begin{figure}
  \centering
  {\includegraphics[width=0.85\textwidth]{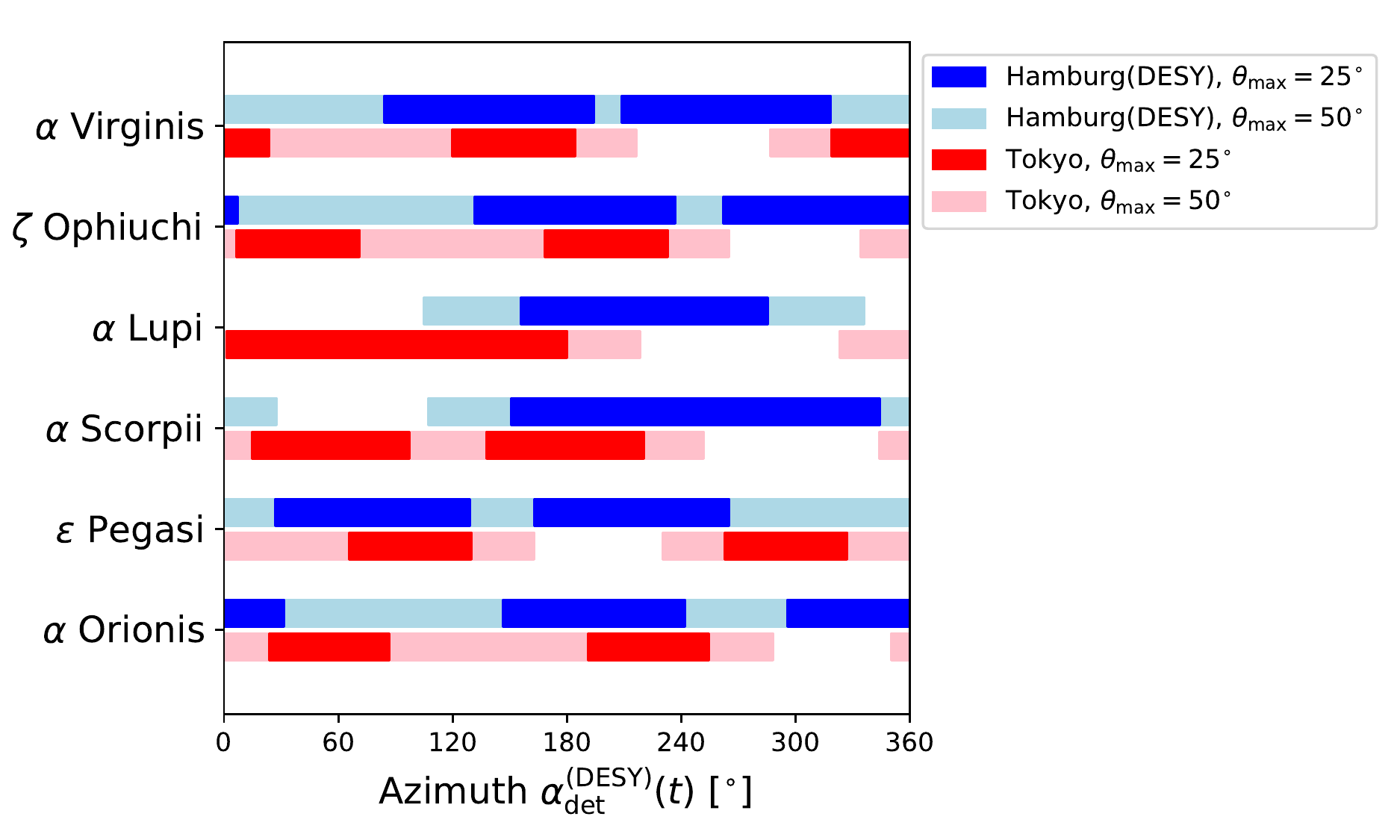}} 
  \caption{
    The observable time of each progenitor at DESY and Tokyo ($\delta_{\rm det} = 35.7^\circ$) in blue and red, respectively, with dark (light) bands corresponding to $\theta_{\rm max} = 25^\circ$ ($50^\circ$). The longitudes of these two points differ by $\simeq 130^\circ$.
  }
  \label{fig:gantt_chart}
  \end{figure}

The effective observational time fractions for these progenitors can be much improved if we simultaneously operate two SNscopes at different observation points or if the maximum elevation $\theta_{\rm max}$ is increased. To see this, in Fig.~\ref{fig:gantt_chart}, we show the observable time of each progenitor at DESY and Tokyo ($\delta_{\rm det} = 35.7^\circ$) in blue and red, respectively, with dark (light) bands corresponding to $\theta_{\rm max} = 25^\circ$ ($50^\circ$). The longitudes of these two points differ by $\simeq 130^\circ$. As we see, the axion SNscopes at these two positions complement each other. The extension of the maximum elevation also has significant impact on the observational time fraction; for instance, if IAXO has $\theta_{\rm max} = 50^\circ$, it can always be directed at $\alpha$ Virginis, $\zeta$ Ophiuchi, $\epsilon$ Pegasi, and $\alpha$ Orionis. Moreover, if both of the SNscopes have $\theta_{\rm max} = 50^\circ$, then all of the progenitor candidates in Table~\ref{tab:nearbystars} can always be observed by at least one SNscope.

\subsection{Event number}
\label{sec:eventnumber}
	In the first $10$ seconds after the bounce of a core-collapse SN, the axion can be abundantly produced and eventually reach the SNscope on Earth. The dominant production process is the nucleon-nucleon bremsstrahlung in the proto-neutron star,
	\begin{equation}
	N\,N' \rightarrow N\,N'+a ~,
	\end{equation}
	where $N$, $N'$ can be either proton or neutron. This process was originally calculated in the one-pion exchange\,(OPE) approximation~\cite{Iwamoto:1984ir,Turner:1988bt,Brinkmann:1988vi}. However, it was realized that various effects in the nuclear medium can reduce the axion emissivity with respect to the OPE evaluation~\cite{Raffelt:1993ix,Hannestad:1997gc,Chang:2018rso,Carenza:2019pxu}. To estimate the event number obtained with the SNscope, we follow the result presented in Ref.~\cite{Carenza:2019pxu}, which considered corrections from
	the non-vanishing mass of the exchanged pion, the $\rho$
	meson exchange, the effective nucleon masses in the medium, and the multiple scattering of nucleons. These effects reduce the axion emissivity from the OPE approximation by about an order of magnitude. The study was done with a simulated $18M_\odot$ {progenitor} and is applicable to heavier SN progenitors with mass $\gtrsim 10M_\odot$, such as the SN1987A. 
	
	Nevertheless, we emphasize that the result in \cite{Carenza:2019pxu} and thus the event number estimated in this subsection should be considered as a guide rather than a precise evaluation. 
	Besides the uncertainty of the nuclear medium effects, the detail composition of the proto-neutron star remains as an open question~\cite{Fischer:2016cyd,Sukhbold:2015wba}.
	The neglected feedback from the axion emission on the SN further alters the axion emission rate, especially when the axion luminosity is close to that of the neutrino. These effects further obscure the prediction of the axion emissivity from an SN. 
	
	We start by showing that the energies of the emitted SN axions are mostly at the
	$\mathcal O$(10)~MeV scale.
	The spectrum of the axions emitted by a unit volume $\dd V$ with temperature $T$ can be expressed as \cite{Raffelt:1993ix,Hannestad:1997gc}
	\begin{equation}
	\frac{{\dd}{ N}_{a}}{{\dd}\omega \dd t \dd V}
	=\frac{1}{4{\pi}^{7/2} f_a^2}
	\left(\frac{g_{A}}{2f_{\pi}}\right)^{4}
	n_{B}\rho\;
	\omega
	\sqrt{m_N T}
	e^{-\omega/T} s(\omega/T)~,
	\label{eq:axion_spectrum_unitV}
	\end{equation}
	where $\omega$ is the energy of the emitted axion, $g_A\simeq 1.26$ is the axial charge, $n_B$ is the baryon number density, and $\rho$ is the mass density of the unit volume.
	$s(x)$ is a dimensionless function that depends on the detailed composition of the proto-neutron star medium and the axion-nucleon couplings $C_N$. With the normalization of Eq.~\eqref{eq:axion_spectrum_unitV}, $s(0)=C_n^2 Y_n^2 +C_p^2 Y_p^2 + \frac{4}{3} Y_n Y_p (3C_+^2 + C_-^2)$ with the OPE approximation and in the limit of massless pion and non-degenerate nucleon. Here, $Y_{n(p)}$ is the fraction of neutron\,(proton) and $C_\pm \equiv (C_n\pm C_p)$\,.
	
\begin{figure}
		\centering
		{\includegraphics[width=0.7\textwidth]{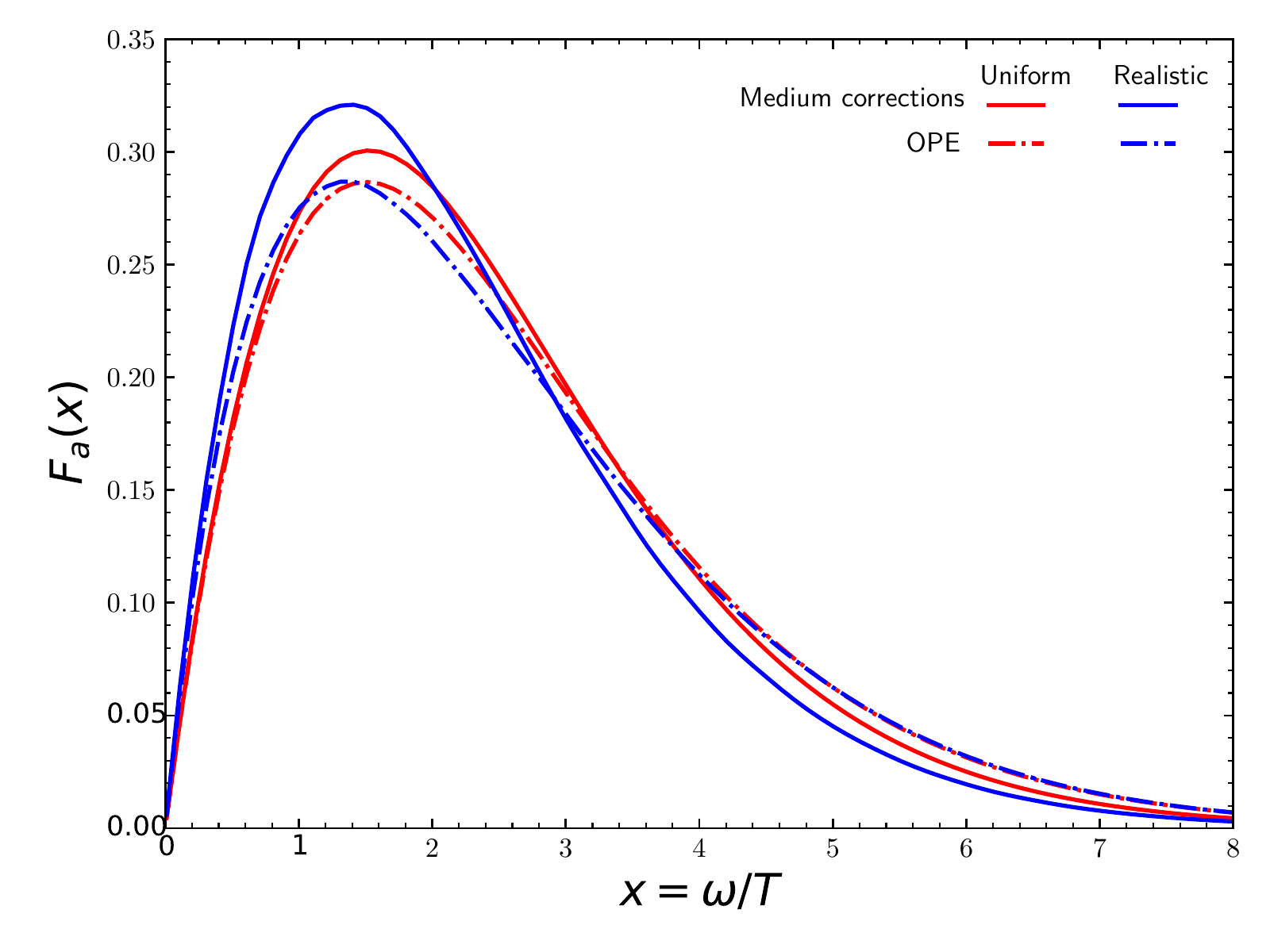}} 
		\caption{The normalized axion spectrum
			$F_a(\omega/T)$ for a unit volume of the SN medium. The dashed and the solid curves represent the results obtained with OPE and with medium corrections respectively. The red curves are plotted for a schematic uniform SN and the blue curves are plotted for a simulated SN~\cite{Carenza:2019pxu}.
		}
		\label{fig:normalized_spectrum}
	\end{figure}

	Even though the exact form of $s(x)$ is highly reliant on the uncertain medium effect~\cite{Carenza:2019pxu}, the shape of the spectrum is only mildly affected. In Fig.~\ref{fig:normalized_spectrum}, we display the scaled axion spectrum
	$F_a(\omega/T)$ from Eq.~\eqref{eq:axion_spectrum_unitV}, which is normalized as $\int\!\dd x\,F_a(x) = 1$, for
	a unit volume of the SN medium.
	The dash-dotted and the solid curves represent 
	results obtained with OPE and with medium corrections described in \cite{Carenza:2019pxu}, respectively. The red curves are plotted for a schematic uniform SN with temperature $T=30$\,MeV, density $\rho=3\times 10^{14}\,{\rm g/cm^3}$, and proton fraction $Y_p=0.3$ {for the KSVZ}. The blue curves are plotted with the result from \cite{Carenza:2019pxu} for a simulated SN at a position of $10$\,km from the center and $1$s post-bounce. We conclude that the average axion energy is $\langle \omega \rangle\approx (2.2-2.5) \, T$. Over $99.9\%$ of the emitted axions are more energetic than $1$~MeV because of the high SN temperature, and typically have energy of ${\cal O}(10)$~MeV.
	The momentum dependence of the axion coupling
	with fermions in Eq.~\eqref{eq:axion_nucleon}
	allows the axion to take away large momentum.
	
	Despite the evidently well determined shape of the spectrum, the height of the spectrum remains less understood. After integrating over the axion energy in Eq.~\eqref{eq:axion_spectrum_unitV}, the
	total number of emitted axion scales roughly as $T^{5/2}$ according to dimensional analysis together with the fact that $s(x)$ is almost temperature independent and only a function of $x$. This magnifies the uncertainty from the SN temperature. Moreover, the constituent of the SN and the medium effect can also alter $s(x)$ by an ${\cal O}(1)$ factor; Therefore, we expect roughly an order of magnitude uncertainty in the estimated total number of axion emitted from the SN.  
	
	The axion luminosity of a SN was estimated at the post-bounce time $t_{\rm pb} = 1$\,s in Ref.~\cite{Carenza:2019pxu} as,
	\begin{equation}
	L_{a}\simeq2.42\times10^{70}\,{\rm erg\cdot s^{-1}}
	\times \left(\frac{m_N}{f_a}\right)^2
	C_{N,{\rm eff}}^2~,
	\label{eq:axion_luminosity}
	\end{equation}
	with 
	\begin{equation}
	C_{N,{\rm eff}}^2\equiv C_{n}^{2}+0.61C_{p}^{2}+0.53C_{n}C_{p}~.
	\end{equation}
	In the first $10$\,s post-bounce, the temperature of the proto-neutron star only varies between $(20- 40)$~MeV in its densest region of $r\lesssim 10$~km~\cite{Fischer:2016cyd,Carenza:2019pxu}. For a rough estimation, we approximate the SN as an isothermal object with $T
	\approx 30$\,MeV. The average emitted axion energy is $\langle\omega\rangle\approx 2.3T\approx 70$~MeV. The rate of axion emission is thus,
	\begin{equation}
	\dot N_{a}\simeq
	\frac{L_{a}}{\langle\omega\rangle}
	\simeq
	2.2\times10^{74}\,\ {\rm s^{-1}}
	\times \left(\frac{m_N}{f_a}\right)^2
	C_{N,{\rm eff}}^2~.
	\label{eq:axion_rate}
	\end{equation}
	
	The probability of an axion converted into a photon inside a helioscope
	is~\cite{Sikivie:1983ip,Sikivie:1985yu,Raffelt:1987im}
	\begin{equation}
	P=\frac{1}{4}\left(g_{a\gamma\gamma}BL\right)^{2}\left(\frac{\sin\left(qL/2\right)}{qL/2}\right)^{2}
	~,
	\end{equation}
	where $q=|m_{a}^{2}-m_{\gamma}^{2}|/(2\omega)$ and $m_{\gamma}$ is the plasma
	mass of the photon.
	For maximum conversion
	of axion to photon, $qL\ll1$ and $P\approx \left(g_{a\gamma\gamma}BL\right)^{2}/4$.
	For an evacuated helioscope, $m_{\gamma}=0$ and the maximum axion
	mass that can be probed efficiently is 
	$m_{a}\lesssim\sqrt{2\omega/L}$. The typical energy of solar axion produced by the
	Primakoff process is $4$\,keV and the helioscopes
	in Table~\ref{tab:helioscopes} starts to lose sensitivity when $m_{a}\gtrsim 0.01~{\rm eV}$.%
	\footnote{The sensitivity to heavy axion can be enhanced by filling the magnet beam
	pipes with buffer gas at various pressure so that the plasma frequency $m_\gamma$ matches $m_{a}$. However, the enhanced sensitivity is worse than that to light axions because of the declined data taking time. CAST has filled its conversion pipe with $^{4}{\rm He}$~\cite{Arik:2008mq,Arik:2015cjv} and  $^{3}{\rm He}$~\cite{Arik:2013nya,Arik:2011rx} to extend its reach to $m_{a}\sim 1~{\rm eV}$.
	IAXO also considers a buffer gas phase as its later improvement to enhance its sensitivity to $m_a= (0.01 - 0.25)$~eV~\cite{Armengaud:2014gea,Armengaud:2019uso,Irastorza:2018dyq}.}
	In contrast, the average SN axion energy is around $70$\,MeV and allows the SNscope to probe heavier axions up to $1$\,eV.

	To summarize, the number of events in the SNscope {for 
	$m_{a}\lesssim\sqrt{2\langle\omega\rangle/L}$} is estimated as
	\begin{subequations}
		\begin{align}
		N &\approx P \dot{N}_{a}\ \frac{A}{4\pi d^{2}} \Delta t \\
		&\approx 1.0\times
		\left( \frac{A}{2.3\,{\rm m^2}} \right)
		\left( \frac{B}{2.5\,{\rm T}} \right)^2
		\left( \frac{L}{20\,{\rm m}} \right)^2
		\times
		\left( \frac{150\,{\rm pc}}{d} \right)^2
		\left( \frac{T}{30\,{\rm MeV}} \right)^{5/2}
		\left( \frac{\Delta t}{10\,{\rm s}} \right)
		\nonumber\\
		&\qquad
		\times
		\left( \frac{C_{a\gamma\gamma}}{0.0023} \right)^2 
		\left( \frac{3\times 10^{8}\,{\rm GeV}}{f_a} \right)^4
		\left( \frac{C_{N,{\rm eff}}}{0.37} \right)^2 ~,
		\label{eq:event_num_scaling}
		\end{align}
	\end{subequations}
	{while for $m_{a}\gg\sqrt{2\langle\omega\rangle/L}$, the number of events is further suppressed by a factor of 
	$(qL/2)^{-2}$ due to the loss of coherence between the axion and the photon field.}
	$A$, $B$, and $L$ are the helioscope parameters listed in Table~\ref{tab:helioscopes},
	$d$ is the
	distance to the SN, $\Delta t$ is the time of observation, and $C_{a\gamma\gamma}\equiv g_{a\gamma\gamma}f_a$.
	We assume that the detection efficiency of the $\gamma$-ray detector is almost 100\%. (See also discussion in Sec.~\ref{sec:bg}.)
	We take $\Delta t\simeq10$\,s since the axion flux remains near its peak value
	in the first 10s post bounce~\cite{Fischer:2016cyd,Carenza:2019pxu}. 
	In Eq.~\eqref{eq:event_num_scaling}, we show the scaling of the expected number of events with respect to the properties of the helioscope, the SN and the axion model. In the following discussion, we always assume the SN temperature $T\simeq 30$\,MeV for estimation.

	\begin{figure}
		\centering
		{\includegraphics[width=0.48\textwidth]{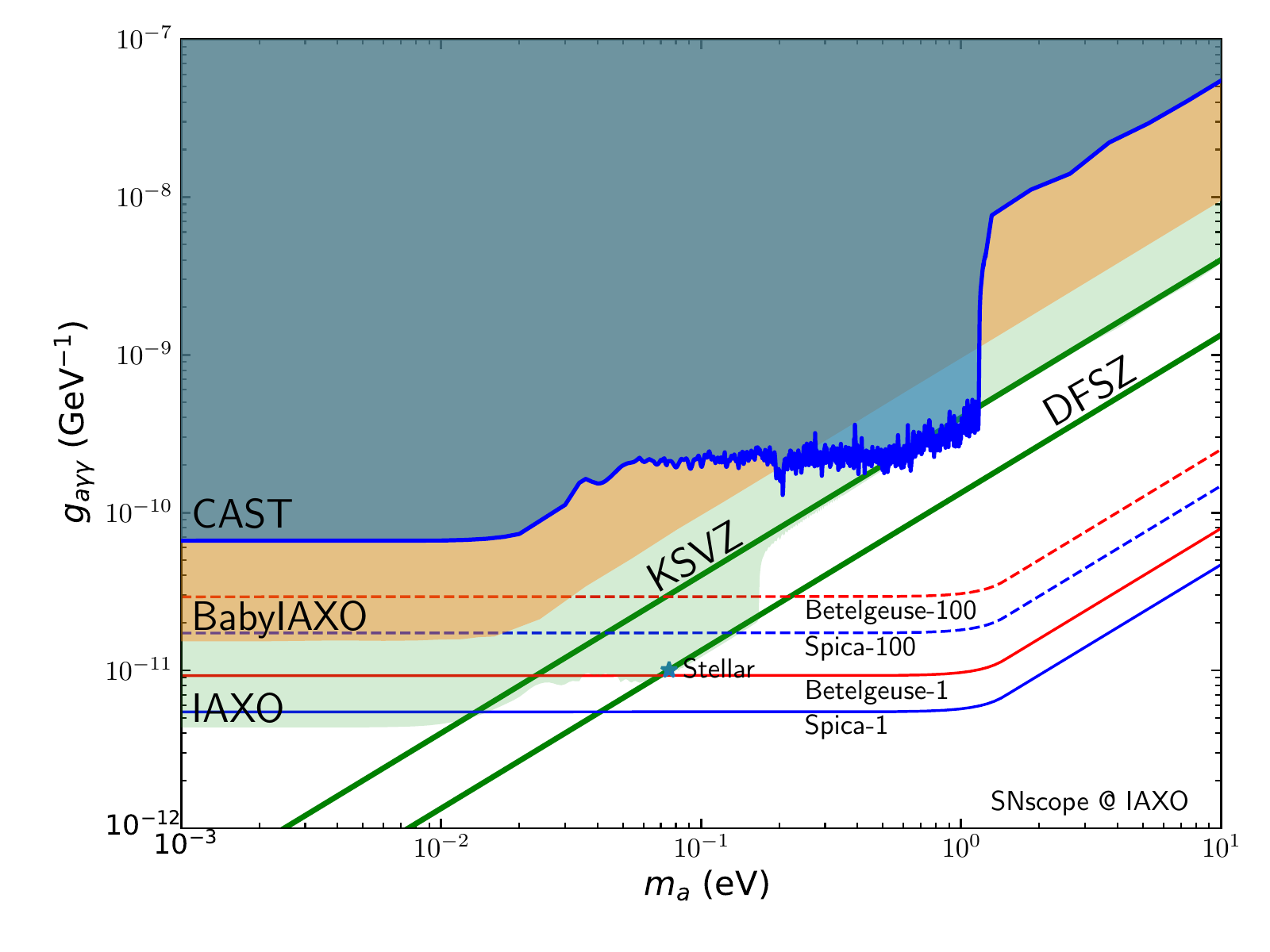}} 
		{\includegraphics[width=0.48\textwidth]{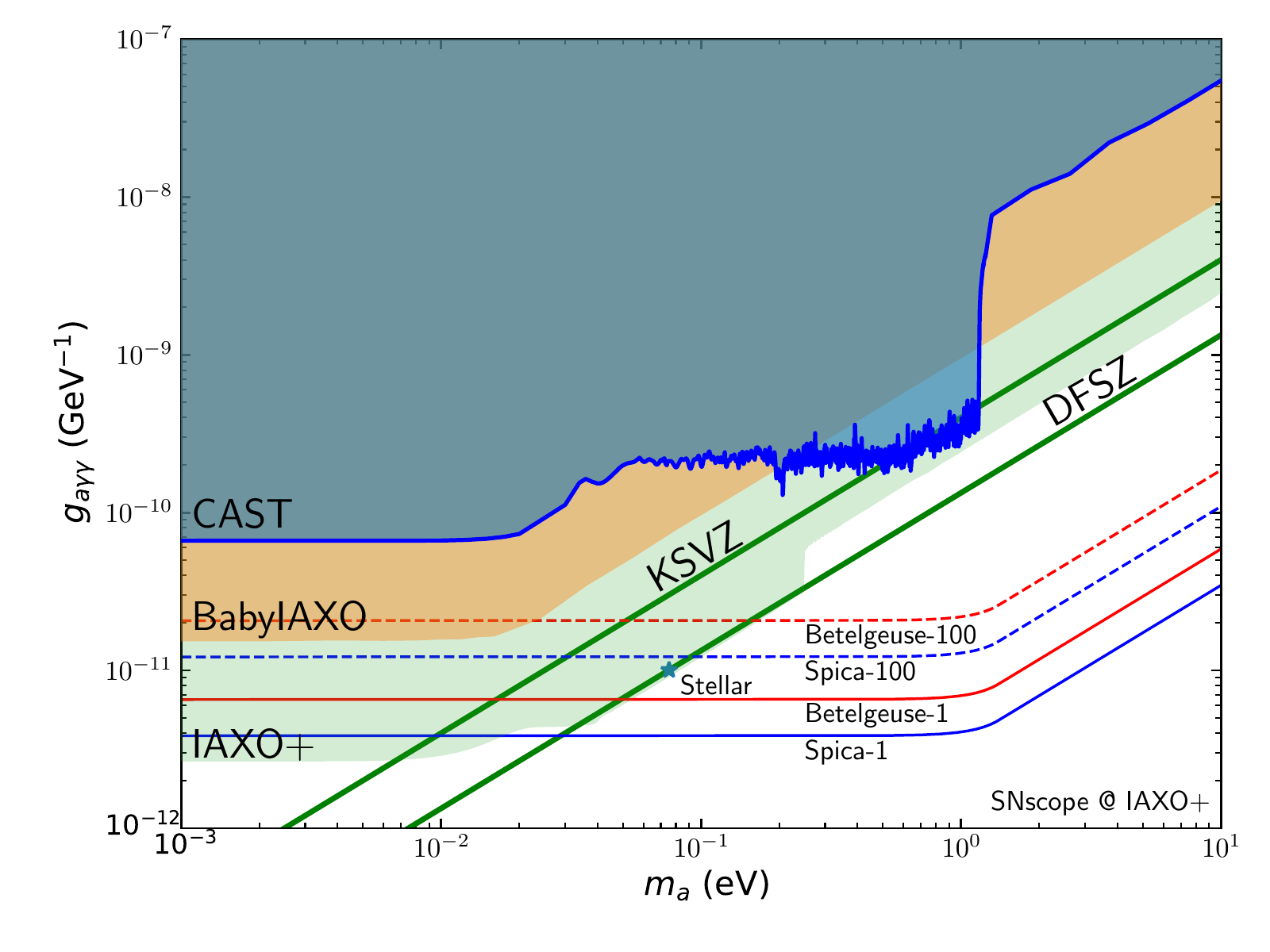}} 
		\caption{The prospect of the SNscope implemented
			at IAXO (left) or IAXO+ (right) helioscope with
			1 or 100 events for the baseline IAXO SNscope. For illustration we take
			Spica and Betelgeuse progenitors in Table \ref{tab:nearbystars}. To obtain the event numbers, we have fixed $C_{N,{\rm eff}}= 0.37$ (KSVZ) and $C_{a\gamma\gamma}\equiv g_{a\gamma\gamma}f_a =\alpha/\pi$. 
		}
		\label{fig:event_contour_gg-ma}
	\end{figure}
	In Fig.~\ref{fig:event_contour_gg-ma}, we plot the contours of 1 and 100 expected events in the $g_{a\gamma\gamma}$-$m_a$ plane for the proposed SNscope implemented at IAXO\,(left panel) or IAXO+\,(right panel).
	The red and the blue contours are plotted respectively for the red supergiant Betelgeuse ($\alpha$ Orionis) and the blue supergiant Spica ($\alpha$ Virginis); the predictions for other progenitors in Table \ref{tab:nearbystars} fall between these two curves. {
	To obtain the event numbers, we have fixed the axion parameters at the values in the KSVZ model, $C_{N,{\rm eff}}=0.37$ and $C_{a\gamma\gamma}\equiv g_{a\gamma\gamma}f_a =\alpha/\pi$, while taking the axion mass $m_a$ as a free parameter. Thus, the event number is proportional to $g_{a\gamma\gamma}^{4}$ in Fig.~\ref{fig:event_contour_gg-ma}. The event number changes by an ${\cal O}(1)$ factor for different axion models; for instance, for the DFSZ model with $E/N=8/3$, the number increases by a factor of 4 ($\tan\beta \ll 1)$ to 14 $(\tan\beta\gg 1)$.}
	The exclusion region by CAST and the projected exclusion regions by IAXO (as well as BabyIAXO and IAXO+) are taken from 
	\cite{Armengaud:2019uso}. The parameter space for the KSVZ and DFSZ axion models are shown with the green lines. The green asterisk marks the ``stellar hint'' of the DFSZ axion with $f_a=7.7\times 10^7\,$GeV and $\tan\beta=0.28$, the best-fit value to the cooling of white dwarfs, red giant branch stars and horizontal branch stars~\cite{Giannotti:2017hny}. For $m_a<10^{-3}$\,eV, the contours remain flat and we ignore this region in the figure. The SNscope starts to lose its sensitivity for $m_a\gtrsim 1$\,eV due to the loss of coherence between the axion and the photon field. Since the energy of the SN axion is much higher than that of the solar axion, the SNscope probes heavier $m_a$ than the helioscopes. This is evident in the region of $m_a\gtrsim 0.05$\,eV for the DFSZ model which is unreachable by the IAXO helioscope in its buffer gas phase.
	 
	\begin{figure}
	\centering
	\includegraphics[width=0.48\textwidth]{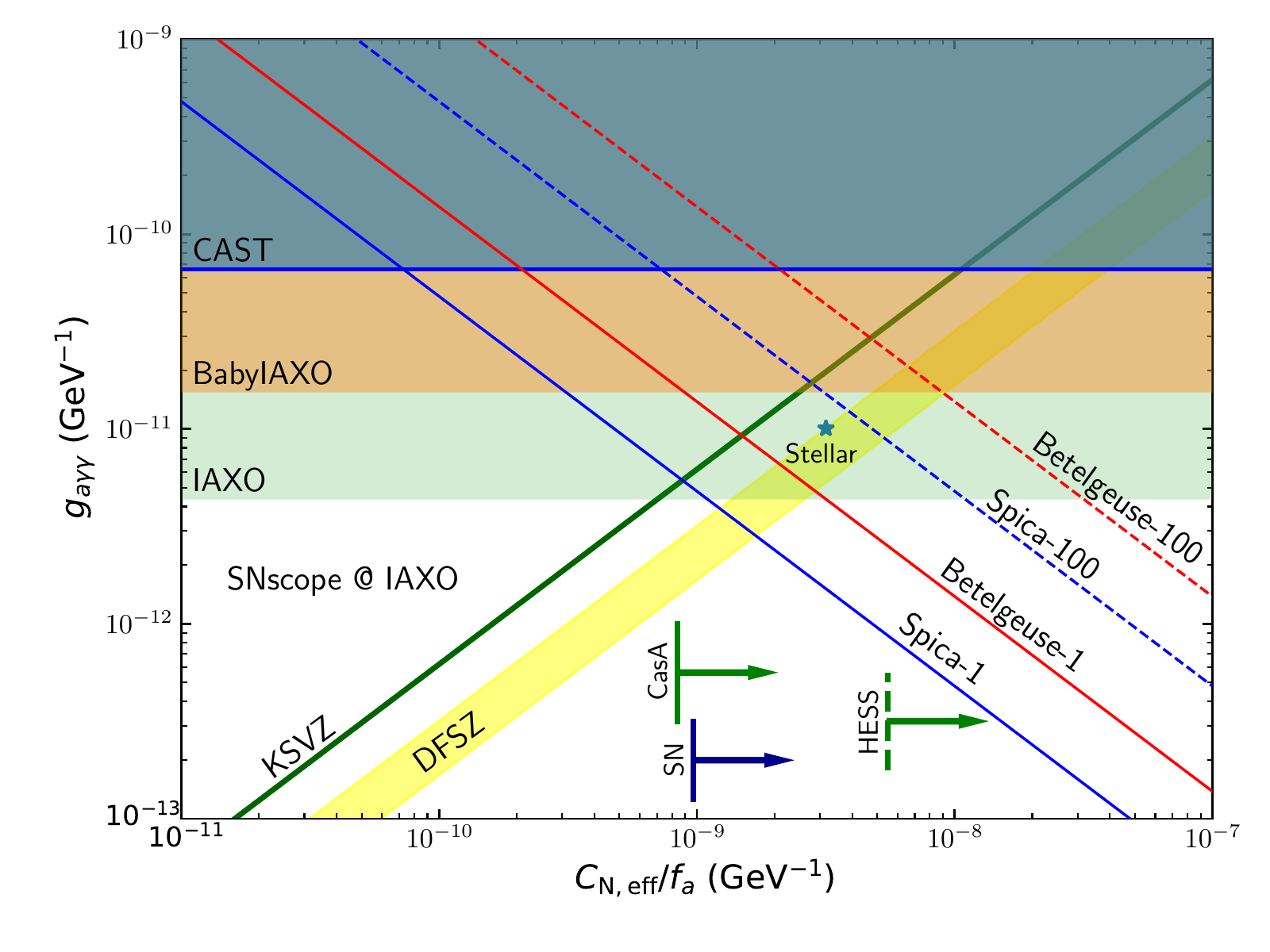}
    \includegraphics[width=0.48\textwidth]{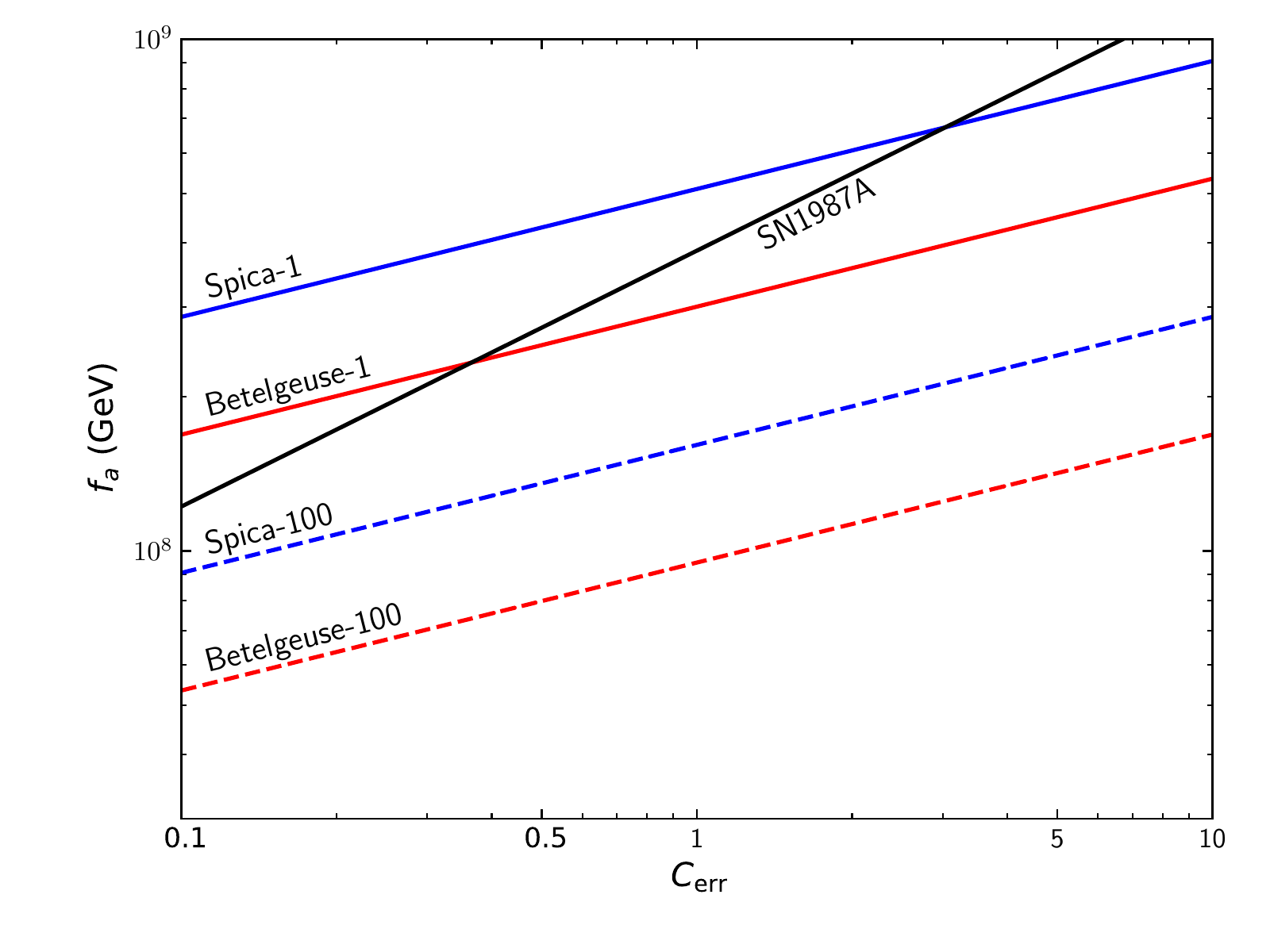}
	\caption{{\it Left panel:} the prospect of the SNscope implemented
		at the IAXO helioscope.
		For illustration we take
		Spica and Betelgeuse progenitors in Table \ref{tab:nearbystars}, and show the contours of 1 (100) event detection by blue and red solid (dashed) lines, respectively.
		We have fixed $m_a = 10^{-3}~\mbox{eV}$; {\em Right panel}: The sensitivity of the SN1987A bound and the 1/100 event contours to the uncertainty of SN axion emission. The latter is parameterized as an overall scaling factor $C_{\rm err}$ to the axion flux.
		\label{fig:event_contour_gg_CN}}
    \end{figure}	
	In the left panel of Fig.~\ref{fig:event_contour_gg_CN}, we plot similar contours as Fig.~\ref{fig:event_contour_gg-ma} but in the plane of 
	$C_{N,\text{eff}}/f_a$ and $g_{a\gamma\gamma}$ and with fixed 
	$m_a = 10^{-3}~\mbox{eV}$ for the baseline IAXO SNscope. {The green line and the yellow band in this figure represent the relation of $C_{N,\text{eff}}/f_a$ and $g_{a\gamma\gamma}$ for the KSVZ and DFSZ axion models, respectively.}
	The projection of the baseline IAXO helioscope covers the parameter space where the QCD axion emitted from a nearby SN gets detected by the IAXO SNscope. If an axion signal is obtained by both the helioscope and the SNscope, it provides valuable structural information of the SN convoluted with the nucleon couplings of the axion. This could be further disentangled in the most optimistic scenario where the event rate in the IAXO helioscope confirms the ``stellar hint'' of the DFSZ axion. 
	
	The dark blue, the green solid and the green dashed arrows in the left panel of Fig.~\ref{fig:event_contour_gg_CN} presents the astrophysical bounds on $C_{N,\text{eff}}/f_a$ for KSVZ axion from the SN1987A~\cite{Carenza:2019pxu}, the neutron star in Cas A\,\cite{Hamaguchi:2018oqw} and the neutron star HESS J1731-347\,\cite{Beznogov:2018fda}. These bounds should be regarded as indicators rather than strict constraints since the detail structure and the processes in these stars are not well understood. The cooling processes of neutron stars depend heavily on property of their envelope and the nucleon superfluid in their core. This could lead to ${\cal O}(1)$ uncertainty in the bound. For the SN1987A, a recent study\,\cite{Bar:2019ifz} argues that the observed neutrino events might have come from an accretion disk that would not be cooled by axions and this invalidates the bound all together. Even if the SN neutrinos and axions are emitted from the proto-neutron star as in the conventional proposal, the medium effect could still decrease the axion flux~\cite{Raffelt:1993ix,Hannestad:1997gc,Chang:2018rso,Carenza:2019pxu}. However, this effect both relaxes the SN1987A bound and reduces the expected number of events in the SNscope. We parameterize the uncertainty of the SN axion flux by an overall scaling factor $C_{\rm err}$, so that the ``true'' axion luminosity and emission rate are parameterized by $\tilde{L}_a= C_{\rm err} L_a$ and $\dot{\tilde{N}}_a= C_{\rm err} \dot{N}_a$, with $L_a$ and $\dot{N}_a$ given by Eq.~\eqref{eq:axion_luminosity} and Eq.~\eqref{eq:axion_rate} respectively. The SN1987A bound is then set by
	$\tilde{L}_a\lesssim L_\nu \simeq 2\times 10^{52}\,{\rm erg}\cdot{\rm s}^{-1}$.
	For a given axion model, the bound on $f_a$ scales as $C_{\rm err}^{1/2}$ while the $f_a$ required to observe one event in the SNscope scales as $C_{\rm err}^{1/4}$. This is demonstrated for the KSVZ model in the right panel of Fig.~\ref{fig:event_contour_gg_CN}. There, the black line depicts the sensitivity of the SN1987A bound on $f_a$ to $C_{\rm err}$. The red and the blue solid (dashed) lines are drawn for Spica and Betelgeuse respectively, and represent the $f_a$ needed to generate 1\,(100) events in the SNscope with varying $C_{\rm err}$. For $C_{\rm err}=0.1$--0.3, the SN1987A observation would be consistent with observing ${\cal O}(10)$ axions from Spica or ${\cal O}(1)$ axions from Betelgeuse with SNscope. Therefore, the observation of axion from a nearby SN is fully plausible even if the SN1987A neutrino was emitted with the conventional mechanism from the proto-neutron star. In any case, the SNscope is a great tool to reserve given the rarity of nearby SN.

\subsection{Background estimate}
\label{sec:bg}

\begin{figure}
  \centering
  {\includegraphics[width=0.55\textwidth]{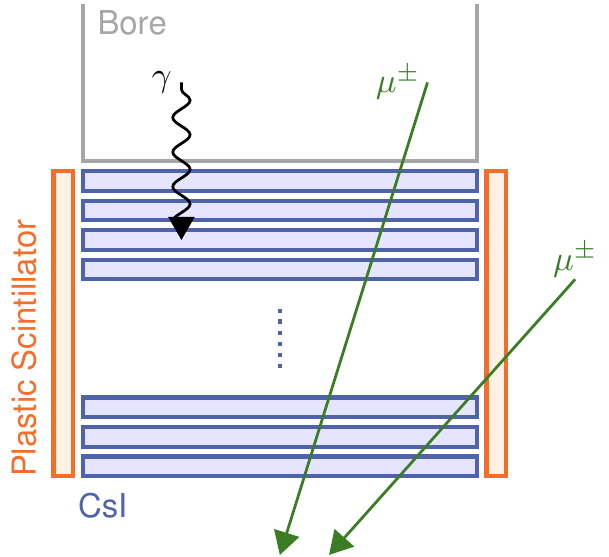}} 
  \caption{
    A design for the gamma-ray detector installed on the axion SNscope. The blue and orange rectangles represent CsI and plastic scintillators, respectively. 
  }
  \label{fig:gam_detector}
  \end{figure}

We now discuss potential background of the axion SNscope. To that end, we consider a concrete setup for the gamma-ray detector, which is illustrated in Fig.~\ref{fig:gam_detector}. We use a set of CsI scintillators (indicated by the blue rectangles) as a gamma-ray detector and pile them up onto the end side of the bore. The side of the detector is surrounded by plastic scintillators (indicated by the orange rectangles), which are used to veto background muons. These scintillators are connected to photosensors to read out scintillation signals. The number of layers of the CsI scintillators will be set large enough so that the photons converted from SN axions deposit most of their energy inside the detector. 

The background for SN axion events in this detector is dominantly caused by cosmic-ray muons. The intensity of cosmic-ray muons at sea level is $\sim 1 ~{\rm cm}^{-2} \cdot {\rm min}^{-1}$~\cite{Tanabashi:2018oca}, and thus for a detector of an ${\cal O}(1)$-meter scale, such as IAXO, we expect $\sim {\cal O}(10^3)$ muons in $\sim 10$ seconds, which is the typical duration of SN axion burst. It is, therefore, crucial to assure that the detector is able to reject cosmic-ray muons with high accuracy. Hereafter, we focus on muons that pass through the CsI plates perpendicularly, assuming that the plastic scintillators are able to reject muons entering from the side of the detector, as indicated by the right green arrow in Fig.~\ref{fig:gam_detector}, with 100\% efficiency.

\begin{figure}
  \centering
  \subcaptionbox{\label{fig:TotalEnergyDeposit}
  Total Energy deposit.
  }
  {\includegraphics[width=0.49\textwidth]{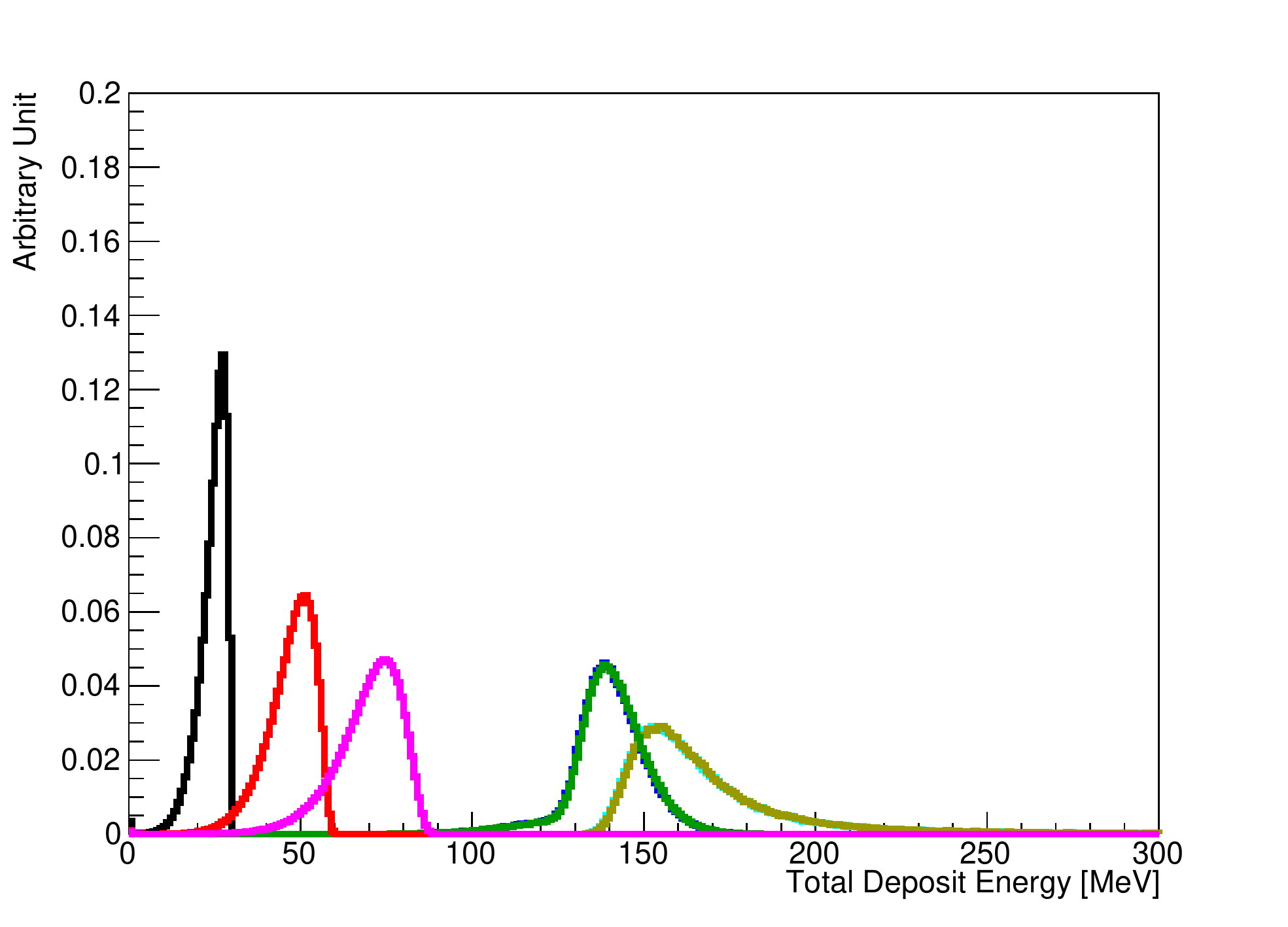}}
  \subcaptionbox{\label{fig:NHitPlate_Gamma90MeV_Muon0.5GeV}
  $N_{\rm hit}$
  }
  { 
  \includegraphics[width=0.49\textwidth]{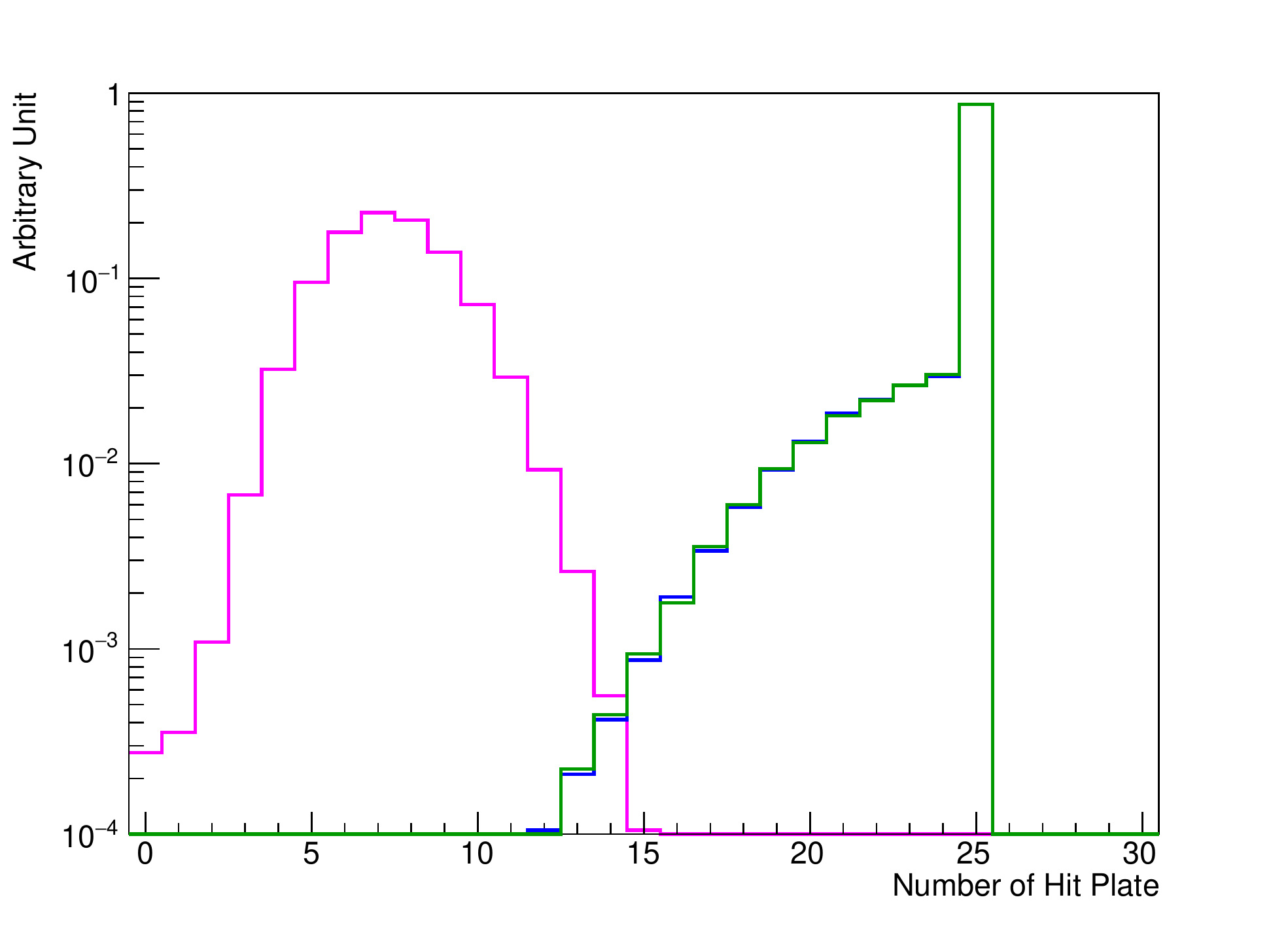}}
  \caption{(a)
     Histograms of the total energy deposit for $30$~MeV (black), $60$~MeV (red), and $90$~MeV (magenta) photons, 0.5~GeV (blue) and 4~GeV (cyan) muons, and 0.5~GeV (green) and 4~GeV (yellow) antimuons;  
    (b) Histograms of the number of plates that receive an energy deposit larger than 1~MeV, $N_{\rm hit}$, for 90~MeV photons (magenta), 0.5~GeV muons (blue), and 0.5~GeV antimuon (green).  
  }
  \label{fig:NHitPlate}
  \end{figure}

To see the prospects for the background muon rejection, we perform Geant4 Monte Carlo simulations~\cite{AGOSTINELLI2003250} using 25 layers of CsI scintillators with a size of $10~{\rm cm} \times 10~{\rm cm} \times 1~{\rm cm}$, which are piled up in parallel and separated by $1$~cm. 
We generate 2 $\times 10^5$ events for 30, 60, and 90 MeV photons and 0.5 GeV and 4 GeV (anti-)muons. All of the particles are injected from above to the center of the top surface of the detector, perpendicularly to the CsI plates.
Here 4 GeV muon corresponds to the muon with mean energy at ground level, and 0.5 GeV muon is chosen as a representative of low energy muon.

Figure~\ref{fig:TotalEnergyDeposit} shows the histograms of the total energy deposit for $30$~MeV (black), $60$~MeV (red), and $90$~MeV (magenta) photons, 0.5~GeV (blue) and 4~GeV (cyan) muons, and 0.5~GeV (green) and 4~GeV (yellow) antimuons. It is found that the photons deposit most of their energy inside the detector; we indicate this feature by the black wavy line in Fig.~\ref{fig:gam_detector}. On the other hand, muons and antimuons lose only a small fraction of their initial energy, and thus penetrate through the detector, as shown by the left green arrow in Fig.~\ref{fig:gam_detector}. The total energy deposited by muons scarcely depends on their initial energy and is almost the same as that for antimuons---and is considerably larger than that for photons. We can use this total energy deposit information to eliminate the muon background events. 

These muons can also be removed by using the information of the number of plates that receive an energy deposit of more than 1~MeV, $N_{\rm hit}$, as shown in Fig.~\ref{fig:NHitPlate_Gamma90MeV_Muon0.5GeV}. For example, if we apply the selection of $N_{\rm hit} \le 12$, an ${\cal O}( 10^{-4})$ reduction of 0.5 GeV muon is achieved while keeping the $\ge 99.5$\% acceptance of the 90 MeV $\gamma$ signal events.  Reconstructing the muon trajectory with segmented CsI scintillators also distinguishes signals and backgrounds. The trajectory reconstruction can be improved if the CsI bars are placed perpendicular to each other for neighboring layers.
The distribution of energy deposits over the CsI plates may 
also be useful to distinguish the photon and muon events, even if both the total energy deposit and $N_{\rm hit}$ are similar in these two cases. 

A subdominant contribution to the background may be provided by nuclear reactions in the detector material. In particular, neutron captures on nuclei in the detector material emit $\gamma$-rays, which can mimic the signal event. It is, however, found that the energies of such $\gamma$-rays are less than 10~MeV (see, \textit{e.g.}, Ref.~\cite{Iida:2020acx}), and thus these $\gamma$-rays can be discriminated from the photons converted from SN axions, as can be seen from Fig.~\ref{fig:TotalEnergyDeposit}. 

All in all, we conclude that it is feasible to sufficiently reduce the number of background events in the axion SNscope, with a realistic setup for the $\gamma$-ray detector. We, therefore, strongly encourage those who work on axion helioscopes to seriously consider the additional installation of such a $\gamma$-ray detector, in order to equip them with the ability of axion SNscopes.

\section{Conclusion}
\label{sec:conclusion}

If a SN explosion occurs within a few 100 pc from the Earth, it would be a once in a lifetime opportunity for directly detecting SN axions. Together with the high statistics measurements of the SN neutrinos expected for such a nearby SN, the (non-)detection of SN axions will provide valuable information about both the SN and the axion physics. 
In this paper, we have studied the prospect of directly detecting the SN axions with an axion helioscope equipped with a $\gamma$-ray detector, which we call SNscope.
If the $\gamma$-ray detector is installed at the opposite end to the X-ray detector for the solar axion, the experiment still functions as an axion helioscope during the normal operation time, whereas it can work as an axion SNscope once the pre-SN neutrinos alert is received.

There are several nearby SN candidates within hundreds of parsecs. With the help of the pre-SN neutrino events, we may be able to narrow the SN candidates down to a few, or may even uniquely identify the one. An axion SNscope can then be pointed to the (one of) the SN candidate(s) in advance. We have shown that the observation time fraction, i.e., the time fraction when the SNscope can be targeted to a SN progenitor within a day, is larger than 50\% for most of the nearby SN candidates, for the IAXO helioscope located at DESY and with a maximal elevation of $\pm \theta_{\rm max} = \pm 25^\circ$. The effective observational time fraction significantly increases if two or more SNscopes at different places operate at the same time and/or the maximal elevation $\theta_{\rm max}$ is increased.

The expected number of SN axions detected at a SNscope is estimated in Sec.~\ref{sec:eventnumber}, and summarized in Figs.~\ref{fig:event_contour_gg-ma} and \ref{fig:event_contour_gg_CN}. If a SN explosion occurs within 250 pc, ${\cal O}$(1--100) events can be detected for an axion decay constant $f_a\simeq (0.9-5)\times 10^8$ GeV taking the KSVZ axion as an example.
We have also discussed the background events, in particular the muon background. By combining the veto by the 
plastic scintillators around the detector, the total energy deposit, and the number of plates {with energy deposit}, it is feasible that we can sufficiently reduce the number of background events at the axion SNscope. 

{
In summary, our study suggests that the SN axion detection is a realistic and promising option for future axion helioscopes and thus is worth being considered seriously. Further studies, including detailed simulations dedicated to each experimental setup with a concrete design for a $\gamma$-ray detector, are highly motivated, which we leave for future work.}

\section*{Acknowledgments}

We are grateful to the organizers of the workshop, ``Revealing the history of the universe with underground particle and nuclear research 2019'' at the Tohoku University, where this project has been initiated.
This work is supported in part by the Grant-in-Aid for Innovative Areas (No.19H05810 [KH], No.19H05802 [KH], No.18H05542 [NN]), Scientific Research B (No.20H01897 [KH and NN]), and Young Scientists B (No.17K14270 [NN]). 
SFG is grateful to the Double First Class start-up
fund (WF220442604) provided by Tsung-Dao Lee Institute
and Shanghai Jiao Tong University.
JZ is supported in part by the NSF of China
(No.11675086 and 11835005).

{\small 
\bibliographystyle{utphysmod}
\bibliography{ref}
}

\end{document}